\RequirePackage[hyphens]{url}
\documentclass{aa}  

\usepackage[toc,page]{appendix}
\usepackage{graphicx}
\usepackage{caption}
\DeclareCaptionLabelFormat{bold}{\textbf{#1 #2}}
\usepackage{subcaption}
\usepackage{txfonts}
\usepackage{natbib}
\usepackage{amsmath}
\usepackage{amssymb}
\usepackage{url}
\usepackage{float}
\PassOptionsToPackage{hyphens}
{url}\usepackage[colorlinks=true,urlcolor=blue,citecolor=blue,linkcolor=blue]{hyperref}
\usepackage{xcolor}
\usepackage{array, multirow}
\usepackage{multicol}
\usepackage{threeparttable}
\usepackage{longtable}
\usepackage{booktabs}
\usepackage{pdfpages}
\usepackage{placeins}
\usepackage{tabularx}
\usepackage{siunitx}
\makeatletter
    \newcommand\ph{$\phantom{1}$}
    \newcommand{\vast}{\bBigg@{3}}
    \newcommand{\Vast}{\bBigg@{3.5}}
    \newcommand{\vastt}{\bBigg@{11}}
    \newcommand{\Vastt}{\bBigg@{13}}
\makeatother

\usepackage{orcidlink}
\usepackage{soul}
\usepackage[normalem]{ulem}

\def\specchar#1{{\sc{#1}}}    
\def\specand{\,\&\,}          
\def\Halpha{\mbox{H\hspace{0.1ex}$\alpha$}}
\def\Hbeta{\mbox{H\hspace{0.1ex}$\beta$}}

\def\CaIIK{\mbox{Ca\,\specchar{ii}\,\,K}}
\def\CaIIHK{\mbox{Ca\,\specchar{ii}\,\,H{\specand}K}}

\definecolor{indiagreen}{rgb}{0.07, 0.53, 0.03}

\defcitealias{neckel1987spectral}{KPN87}
\defcitealias{Wallace2011ApJS..195....6W}{KPW11}
\defcitealias{Meftah2023RemS...15.3560M}{HRS23}
\defcitealias{Reiners2016}{IAG16}
\defcitealias{Strassmeier2018A&A...612A..44S}{PEP18}
\defcitealias{Ellwarth2023}{IAG23}
\defcitealias{Molaro2013A&A...560A..61M}{LFC13}

\begin{document} 

   \title{Solar flux atlases}

   \subtitle{The new HARPS-N quiet Sun benchmark and continuum
normalisation of the Ca II H \& K lines  }

   \author{
   F. Hanassi-Savari\inst{1,2}\orcidlink{0009-0009-8498-8581}
   \and A.G.M.\ Pietrow\inst{1,3}\orcidlink{0000-0002-0484-7634}
   \and M.K.\ Druett\inst{2,3}\orcidlink{0000-0001-9845-266X} 
   \and M. Cretignier\inst{4}
   \orcidlink{0000-0002-9911-2285}
   \and M. Ellwarth\inst{5,6,7}
   \orcidlink{0009-0007-0922-7315}
   }

   \institute{\inst{1}Leibniz-Institut für Astrophysik Potsdam (AIP), An der Sternwarte 16, 14482 Potsdam, Germany\\
   \inst{2}Plasma Dynamics Group, School of Electrical and Electronic Engineering, University of Sheffield, Sheffield, S1 3JD, UK\\
   \inst{3}Centre for Mathematical Plasma Astrophysics, KU Leuven, Celestijnenlaan 200B, B-3001 Leuven, Belgium\\
   \inst{4}Department of Physics, University of Oxford, OX13RH Oxford, United Kingdom \\
   \inst{5}Institut für Astrophysik und Geophysik, Georg-August Universität Göttingen, Friedrich-Hund-Platz 1, 37077 Göttingen, Germany\\
   \inst{6}Lowell Observatory, 1400 W. Mars Hill Road, Flagstaff, AZ 86001, USA\\
   \inst{7}Department of Astronomy and Planetary Science, Northern Arizona University, PO Box 6010, Flagstaff, AZ 86011 USA\\
   \email{fhanassi-savari1@sheffield.ac.uk}\\
   }

   \date{Submitted: \today}
  \abstract
  {Solar flux atlases observe the spatially integrated light from the Sun, treating it as a star. They are fundamental tools for gaining insight into the composition of the Sun and other stars. They are utilised as reference material for a wide range of solar applications such as stellar chemical abundances, atmospheric physics, stellar activity, and radial velocity signals.} 
  {We provide a detailed comparison of solar activity reported in some of the well-known solar atlases against the new High Accuracy Radial velocity Planet Searcher for the Northern hemisphere (HARPS-N) Quiet Sun (QS) and Measured Activity (MA) atlases published, for the first time, in this work.} 
  {Ten of the widely used individual spectral lines from each flux atlas were selected to compare solar activity based on three methods: 1) equivalent widths; 2) a novel activity measure  introduced in
this work and referred to as the activity number; and 3) bisectors and radial velocity.}
  {The significantly smaller activity levels measured in the MA atlas, compared to the other atlases, relative to the QS atlas, underscores the dominance of instrumental effects over solar activity in their impact on spectral lines, which cannot be corrected through simple line convolution to match resolutions of other atlases. Additionally, our investigation unexpectedly revealed a substantial intensity shift in the \CaIIHK{} lines of other atlases compared to our HARPS-N atlases, likely resulting from the assumptions made when applying normalisation techniques for the early Kitt Peak atlases.}
  {With an average spot number of zero, our QS atlas is well suited to serve as an absolute benchmark atlas representative of solar minimum for the visible spectrum, which other atlases can be compared against. Our recommendations  going forward include: 1) the publication of a detailed log along with the observations to include exact dates and indications of solar activity; and 2) given the dominance of instrumental effects over variations caused by activity, quiet Sun reference atlases must be constructed with the same instruments to ensure high precision.}   

\keywords{Atlases, Sun: activity, Line: profiles, Methods: data analysis}
   \maketitle
\section{Introduction} \label{sec:intro}
Solar atlases, or solar reference spectra, are comprehensive high-resolution collections of data drawn from integrated scans across a wide wavelength range of the solar spectrum. They play a crucial role in numerous branches of both solar and stellar research because the Sun is bright enough to allow for high signal-to-noise (S/N) observations for relatively short integration times. 
Such atlases serve as a primary reference for a wide range of stellar applications. They are particularly useful for calibration of observed spectra \citep[e.g.][]{Lofdahl2021A&A...653A..68L}. Additionally, they help estimate physical processes such as radial velocity (RV; \citep{Trifonov20, AlMoulla2024, Lakeland2024} and convection \citep[e.g.][]{Meunier2010A&A...519A..66M, 2023A&A...680A..62E, Pietrow23}. These atlases are also crucial for understanding the impact of active regions such as spots, plages, and network \citep{Cretignier24,Pietrow2023Nessi}. Furthermore, they help analyse variations introduced by activity cycles \citep{Maunder04, Willamo20} and offer indications of what the solar disc may look like in a given wavelength \citep{VaradiNagy2025}. There are two main kinds of solar atlases, described below. 
\begin{enumerate}
    \item Disc centre (or intensity or photospheric\footnote{The terms disc centre and photospheric are sometimes used interchangeably in the literature which, we believe to be confusing. Disk centre is the accurate terminology for this type of atlas.}) atlases, which represent the average spectrum at the centre of the solar disc. Thus, they do not take into account the curvature and limb darkening of the star. These atlases are frequently used to make comparisons with synthetic spectra that are generated from plane-parallel models \citep{Pereira2013A&A...554A.118P}, as well as for the calibration of data that do not conserve the flux \citep{Beck2011A&A...535A.129B, Lofdahl2021A&A...653A..68L}. Some of the most used disc-centre atlases include: \citet{Liege1973apds.book.....D}, \cite{Neckel1984SoPh...90..205N} and \citet{Wallace1998assp.book.....W}.

    \item Disc-integrated (or flux) atlases, which are typically made by observing a defocussed Sun; for instance, multiple Kitt Peak Fourier Transform Spectrograph (FTS) atlases \citep{Kurucz1984sfat.book.....K, Wallace2011ApJS..195....6W}. Unlike the disc-centre atlas, this type of atlas treats the Sun as if it were a distant star, capturing subtle, but increasingly significant effects such as limb darkening \citep{Ellwarth2023}, differential rotation \citep{CollierCameron2019MNRAS.487.1082C}, and convective blueshifts \citep{Sheminova2022KPCB...38...83S, Dravins2008A&A...492..199D, Gray2009ApJ...697.1032G}. Some commonly used atlases of this were compiled by \citet{Beckers1976hrsa.book.....B}, \citet{Kurucz1984sfat.book.....K}, \citet{Wallace2011ApJS..195....6W}, and \citet{Reiners2016}.
\end{enumerate}
However, the process of observing solar spectra presents numerous challenges that can contaminate or bias the data. Ground-based observations for both types of atlases suffer from Earth atmospheric (or telluric) contamination, as well as instrumental effects. Disk-integrated atlases are also sensitive to solar activity, which affects the line depths and shapes of all spectral lines \citep{Basri(1989),Thompson2017MNRAS.468L..16T,Cretignier(2020a)}. This effect is strongest in chromospheric lines \citep{Livingston2007ApJ...657.1137L, Pietrow23}. Moreover, previous solar atlases have not systematically accounted for solar activity, usually assuming that it cannot significantly influence the spectra \citep[e.g.][]{Doerr2016, Wallace2011ApJS..195....6W} and potentially introducing biases and inaccuracies into their spectral profiles.

To detect Earth-mass planets orbiting in the habitable zones of Sun-like stars, RV sensitivities of around \text{10~cm~s$^{-1}$} are required. As \citet{Fischer2016PASP..128f6001F} highlighted, from an instrumental-precision perspective, projects that leverage laser comb technologies for their wavelength calibrations are expected to achieve this target with current or next-generation spectrometers, albeit for a limited range of the visible spectrum \citep{MolaroMonai2012A&A...544A.125M}. However, reaching this precision alone does not translate into concrete detections of Earth-like planets. The biggest obstacles come from the intrinsic stellar variations, which are manifest in different forms, including magnetic activity in the form of spots and plages, granulation, photospheric rotations (sometimes known as stellar jitter or noise), activity cycles, and so on \citep[see][for a review]{Meunier2021}. The Sun, due to its proximity, is currently the only star where these phenomena can be studied and monitored in detail. Studies such as \citet{Meunier2010A&A...512A..39M}, \citet{Lovis2011}, \citet{Dumusque2011A&A...527A..82D}, \citet{Haywood2016MNRAS.457.3637H},  \citet{Luhn2020AJ....159..235L}, and \citet{Pietrow2024A&A...682A..46P} have indicated that stellar activity can produce RV signals ranging from tens of \text{cm~s~$^{-1}$} up to tens of \text{m~s$^{-1}$}. The introduction of new innovative RV mitigation and detection techniques will be pivotal for breaking the barriers towards the elusive \text{10~cm~s$^{-1}$} sensitivities. 

These challenges highlight the need for an absolute reference spectrum that accurately represents the Sun under quiet conditions. Such a reference spectrum serves as a crucial benchmark for comparing, evaluating, and calibrating other solar atlases. This paper aims to introduce the HARPS-N Quiet Sun (QS) atlas\footnote{Made available with this paper.} as an absolute reference atlas to represent the solar activity minimum and assess its validity against some of the most widely used solar atlases.

In Section~\ref{sec:observations} we  introduce and summarise our own HARPS-N QS and Measured Activity (MA) atlases, where the seven most cited atlases were used: \citet[][hereafter, \citetalias{Ellwarth2023}]{Ellwarth2023}, \citet[][hereafter, \citetalias{Meftah2023RemS...15.3560M}]{Meftah2023RemS...15.3560M}, \citet[][hereafter, \citetalias{Strassmeier2018A&A...612A..44S}]{Strassmeier2018A&A...612A..44S}, \citet[][hereafter, \citetalias{Molaro2013A&A...560A..61M}]{Molaro2013A&A...560A..61M}, \citet[][hereafter, \citetalias{Reiners2016}]{Reiners2016}, \citet[][hereafter, \citetalias{Wallace2011ApJS..195....6W}]{Wallace2011ApJS..195....6W}, and \citet[][hereafter, \citetalias{neckel1987spectral}]{neckel1987spectral}. In Section~\ref{sec:Methods}, we describe the three methods we used to analyse the activity levels of each atlas. The details of the analysis are discussed in Section~\ref{sec: results}. Finally, we present our concluding remarks in Section~\ref{sec:Conclusion}. 

\section{Observations} \label{sec:observations}
\begin{figure*}[t]
    \centering
    \includegraphics[width=0.99\linewidth]{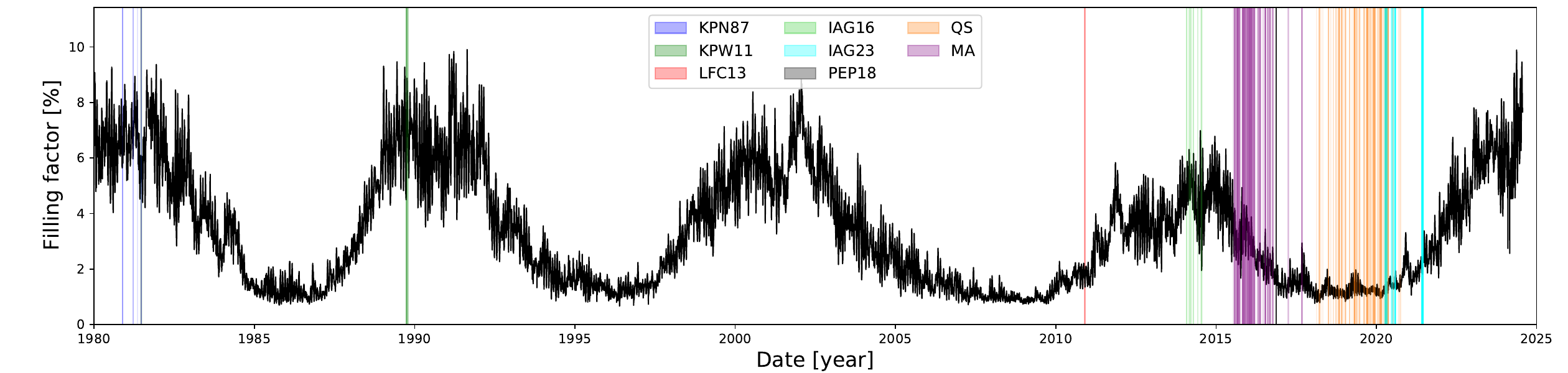}
    \caption{Solar activity over time as measured by the total filling factor of active regions (see main text). The observation time for each of the atlases compared in this study are shown to illustrate the average level of activity. The atlases are  \citetalias{neckel1987spectral}, \citetalias{Wallace2011ApJS..195....6W}, \citetalias{Molaro2013A&A...560A..61M}, \citetalias{Reiners2016}, \citetalias{Strassmeier2018A&A...612A..44S}, HARPS-N's MA \& QS atlases, and \citetalias{Ellwarth2023}, respectively. No information is provided regarding the observing dates of the \citetalias{Meftah2023RemS...15.3560M} atlas.}
    \label{fig:atlassolarcycle}
\end{figure*}

In this section, we first discuss our QS and MA atlases, followed by the seven flux atlases we compared this atlas against. The key parameters of each atlas are summarised in Table~\ref{tab:Atlas_specs}. In addition, URL links are provided for online access to the data from each atlas at the time of publication.
\begin{table}
    \centering
    \captionsetup{justification=raggedright,singlelinecheck=false}
    \caption{Summary of features of the solar flux atlases compared in this paper.} 
    \begin{tabular}{p{1.6cm}p{2.4cm}p{1.3cm}p{1.3cm}p{0.8cm}}
    \hline
    \hline
    Atlas & $\lambda$-range [$\AA$] & $\Delta \lambda$ [$\AA$] & $\lambda/\Delta\lambda$ [$\SI{}{\kilo{}}$] & R [$\SI{}{\kilo{}}$]\\
    \hline
    QS \& MA\textsuperscript{1} & 3900 --- \phantom{0}6834 & 0.0100 & \phantom{0}\SI{500}{} & \ph115\\
    IAG23\textsuperscript{2} & 4200 --- \phantom{0}8000 & 0.001 & \SI{5000}{} & \phantom{0}700\\
    HRS23\textsuperscript{3} & \ph\ph\ph5 --- 44000 & 0.0027 & \SI{2000}{} & 1250\\
    PEPSI\textsuperscript{4} & 3820 --- \phantom{0}9138 & 0.0089 & \phantom{0}\SI{563}{} & \phantom{0}270\\
    IAG16\textsuperscript{5} & 5002 --- 11086 & 0.0084 & \SI{5300}{} & 1000\\
    LFC13\textsuperscript{6} & 4754 --- \phantom{0}5860 & 0.0100 & \phantom{0}\SI{500}{} & \phantom{0}115\\
    KPW11\textsuperscript{7} & 2958 --- \phantom{0}9250 & 0.0038 & \SI{2040}{} & \phantom{0}525\\
    KPN87\textsuperscript{8} & 3290 --- 12500 & 0.0075 & \phantom{0}\SI{417}{} & \phantom{0}435\\
    \hline
    \multicolumn{5}{p{\dimexpr\columnwidth-2\tabcolsep-2\arrayrulewidth}}{
    \parbox{\linewidth}{
        \vspace{0.5em}
        \footnotesize \textsuperscript{1} QS and MA data have been released with this paper.\\
        \footnotesize \textsuperscript{2} The modified version of IAG23 atlas with no telluric extraction have been released with this paper.\\
        \footnotesize \textsuperscript{3} \href{http://bdap.ipsl.fr/voscat_en/solarspectra.html}{\texttt{http://bdap.ipsl.fr/voscat\_en/solarspectra.html}}\\
        \footnotesize \textsuperscript{4} \href{https://pepsi.aip.de/?page_id=549}{\texttt{https://pepsi.aip.de/?page\_id=549}}\\
        \footnotesize \textsuperscript{5} \href{https://zenodo.org/records/3598136}{\texttt{https://zenodo.org/records/3598136}}\\
        \footnotesize \textsuperscript{6} \href{https://cdsarc.cds.unistra.fr/ftp/J/A+A/560/A61/}{\texttt{https://cdsarc.cds.unistra.fr/ftp/J/A+A/560/A61/}}\\
        \footnotesize \textsuperscript{7} \href{https://nispdata.nso.edu/ftp/pub/Wallace_2011_solar_flux_atlas/?_gl=1*x46g27*_ga*ODI0MzA1NjIuMTcxNzA3MzMwMw..*_ga_TW89ZBTY53*MTcxOTc2MzIxMS45LjEuMTcxOTc2MzU3MC42MC4wLjA}{\texttt{https://nispdata.nso.edu/ftp/pub/Wallace2011}}\\
        \footnotesize \textsuperscript{8} \href{ftp://ftp.hs.uni-hamburg.de/pub/outgoing/FTS-Atlas/}{\texttt{ftp://ftp.hs.uni-hamburg.de/pub/outgoing/FTS-Atlas/}}\\
    }
}
    \end{tabular}
    \label{tab:Atlas_specs}
    \vspace{1ex}
    \parbox{\textwidth}{\raggedright \vspace{1ex} \textbf{Notes.} The wavelength ranges (\text{$\lambda$}-ranges) listed here correspond to\\ the data available online for each atlas. The reported wavelength\\ coverage of some atlases are different from the available data\\ sources. Where this is the case we have mentioned this in the\\ respective subsection of each atlas in Section~\ref{sec:observations}. The $\Delta \lambda$ [$\AA$] values\\ are calculated in this work using the nearest two wavelengths to\\ \text{5000~\AA}. Values in the R column represent the authors' average\\ reported resolutions for their respective atlases. Both columns are\\ given in thousands (k).}
\end{table}
In Fig.\ref{fig:atlassolarcycle}, we summarise where all the observations of the different programs were obtained compared with the magnetic cycle as measured by the Mg~II core emission index \citep{Snow2014}. The Mg~II emission Bremen composite index was converted in total filling factor by fitting a linear regression with the total\footnote{Including spot, plages, and network} SDO filling-factor time series extracted in \citet{Cretignier24}. From this figure, we can see that some atlases were actually obtained near solar magnetic cycle maxima; whereas our new QS atlas, as well as \citetalias{Molaro2013A&A...560A..61M}, the \citetalias{Meftah2023RemS...15.3560M} atlas, and the resolved atlas \citetalias{Ellwarth2023} were probing the solar cycle minimum.

\subsection{HARPS-N QS and MA atlases} \label{sec: HARPS-N}
The HARPS-N located at the Telescopio Nazionale Galileo (TNG) on La Palma (Canary Islands) is the complementary project to the similar HARPS instrument installed on the 3.6~m European Southern Observatory (ESO) telescope in Chile \citep{Mayor2003Msngr.114...20M}. HARPS-N is a fibre-fed, high-resolution, \text{\(R = 115{,}000\)}, echelle spectrograph covering the entire visible range. It is extensively described in \citet{Cosentino2012SPIE.8446E..1VC}.

As detailed in \citet{Dumusque2015ApJ...814L..21D} and \citet{Phillips2016SPIE.9912E..6ZP}, a small solar telescope was connected to HARPS-N that has been observing the sun as a star since 2015 \citep{CollierCameron2019MNRAS.487.1082C,Dumusque2021A&A...648A.103D}. The 3~inch lens with 200~mm focal length telescope feeds the HARPS-N spectrograph via a 2~inch diameter integrating sphere connected to an optical fibre. This telescope achieves a high S/N ratio of about 300 at a cadence of 5 minutes. Since July 2015, this solar telescope has been collecting data from the Sun for 4 to 6 hours of every clear day with 5 minute exposures. The 5 minute integration times are used to mitigate any impacts from the 5 minute solar oscillations. 
HARPS-N is known to have stability in wavelength solutions down to Doppler speeds of \text{$\sim$~75~cm~s$^{-1}$} \citep{Dumusque2021A&A...648A.103D}; however, the systematic accuracy in the wavelength solution is not better than dozens of \text{m~s$^{-1}$}, as recently demonstrated for ESPRESSO in \citet{Schmidt2024MNRAS.530.1252S}. 

The HARPS-N wavelength solution is obtained from a combined fit of Thorium-Argon (ThAr) \footnote{\url{https://www.eso.org/sci/facilities/paranal/instruments/uves/tools/tharatlas.html}} lines and Fabry-P\'erot Interferometer measurements. Since the interferometer does not provide an absolute wavelength scale, but only relative ones, the accuracy in wavelength calibration is determined by the measurements in the ThAr lines position, which are known to be accurate locally in the wavelength spectrum down to Doppler speeds of \text{$\sim$~75~m~s$^{-1}$} \citep{Cersullo(2019),Coffinet(2019)}\footnote{It is indeed a coincidence that the stability is \text{$\sim$~75~cm~s$^{-1}$}, but the systematic offset is about \text{$\sim$~75~m~s$^{-1}$}.}. This is the principal limitation in terms of the accuracy of this atlas. It must be noted that a precision of \text{$\sim$~75~cm~s$^{-1}$} in the stability of the wavelength solution means that, on average, two spectra observed at different times will be \text{$\sim$~75~cm~s$^{-1}$} away from one another and not from the ground truth itself. This precision is not indicative of local accuracy of a single observation. 

The spectra used to form the master quiet spectrum (MQS) were selected as the top 10 percent of the most quiet spectra, based on the emission of the \CaIIHK{} lines. This is because these lines have been shown to be the most sensitive to solar activity in the visible \citep{Cretignier24, Pietrow2024A&A...682A..46P, Dineva22}. The epochs of the observations mainly ranges between 24 April 2019 and 12 May 2020. The exact dates can be found in Table~\ref{tab:QS_Spot_Numbers}. This period of time probes the minimum of the solar cycle when no active region was visible on the solar disc and for which only a restricted coverage \text{($\sim$~1\%)} by the magnetic network, in the form of bright points located within the intergranular lanes of the quiet Sun, were present. We used the definition set in the last paragraph of the introduction in \citet{Cretignier24} and refer to Fig. 13 of the same paper for more details.

The processing of the spectra used to form the MQS follows routines developed to achieve the optimal RV precision \citep{Cretignier(2022),Cretignier(2023)}. Specifically, all the spectra are first shifted to the heliocentric rest frame  to remove the signature of the Solar System planets \citep{CollierCameron2019MNRAS.487.1082C} using Jet Propulsion Laboratory (JPL) Horizons ephemeris \citep{Giorgini(1996)} and re-interpolated on a common wavelength grid with a 0.01~\AA\ step. Daily stacking is then performed  to smooth over intra-day timescale phenomena such as oscillations and granulation. 

The daily stacked spectra are then continuum normalised using RASSINE \citep{Cretignier(2020b)}, a public\footnote{\url{https://github.com/MichaelCretignier/Rassine_public}} Python code that uses an alpha shape to fit the upper envelop of the spectra. The continuum normalised spectral time series is then processed by YARARA \citep{Cretignier(2021)}, a post-processing methodology developed to remove instrumental systematics, telluric lines, and stellar activity signatures. YARARA utilises the POLLUX database \citep{Palacios2010}, which contains both the MARCS\footnote{\url{https://web.archive.org/web/20150704031702/http://www.mpa-garching.mpg.de/PUBLICATIONS/DATA/SYNTHSTELLIB/synthetic_stellar_spectra.html}} and ATLAS\footnote{\url{https://www.mpa-garching.mpg.de/PUBLICATIONS/DATA/SYNTHSTELLIB/synthetic_stellar_spectra.html}} models used to scale the continuum in an accurate absolute flux scale. Once the spectra are corrected, the MQS can be obtained by taking the median of the data at each wavelength point from the corrected spectral time series. 

\begin{figure*}
    \centering
    \includegraphics[width=0.99\linewidth, trim={1cm 6cm 1.0cm 6cm},clip]{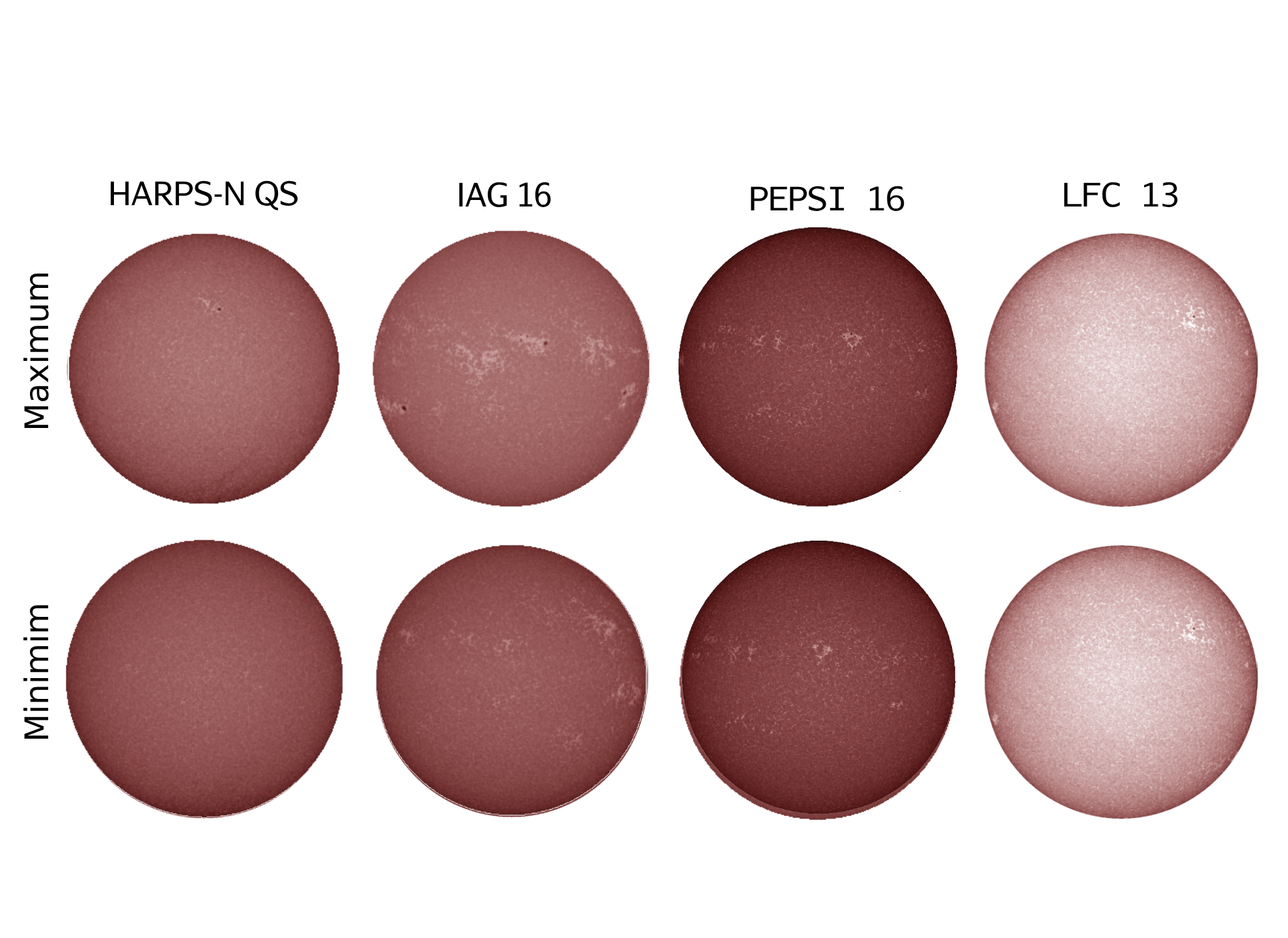}
    \caption{Overview of atlas activity. Each disc on the top row represents the most active, with the highest spot number, day used for said atlas. The bottom row is the same except that it shows the lowest activity level for any given atlas. Data from the Solar Dynamics Observatory Atmospheric Imaging Assembly \citep[SDO/AIA,][]{Lemen2012SoPh..275...17L} 1700~\rm\AA\ filter. Left: HARPS-N quiet Sun atlas,  Reiners IAG flux atlas,  PEPSI 2018 atlas, and  Molaro LFC atlas. We note that the LFC atlas only has one data point, thus, it shows the same data twice. }
    \label{fig:atlascompare}
\end{figure*}

Furthermore, for the purposes of this study, we constructed a second HARPS-N atlas (i.e. MA) as the median of 182 selected moderate activity days of the same solar cycle. As shown in Fig.~\ref{fig:atlassolarcycle}, this atlas represents a medium activity level during the declining phase of solar cycle 24. The exact observation dates for this atlas can be found in Table~\ref{tab:MA_Spot_Numbers}. In the first column of Fig.~\ref{fig:atlascompare}, we show a representation of the activity level seen on the solar disc using the SDO/AIA 1700~\AA\ filter. Each atlas is represented by the most active and the quietest observations from the series. 

This atlas is available on the Strasbourg astronomical Data centre (CDS) connected to the online material for this paper. Both our QS and MA atlases are obtained from the recent data release of a decade-long HARP-N observations \citep{Dumusque2025-nb}.

\subsection{IAG solar CLV atlas:\  \texorpdfstring{\citet{Ellwarth2023}}{Ellwarth et al. 2023}} \label{atlas:ellwarth}
This spatially resolved solar atlas was observed between spring 2020 and summer 2021. The observations were carried out at the Institut für Astrophysik and Geophysics in Göttingen using an FTS in combination with the Vacuum Vertical Telescope (VVT), as in the case of the \citetalias{Reiners2016} integrated Sun atlas. Therefore, the atlases share a resolving power of up to $R = \nu/\Delta \nu \approx 10^6$.  The observations contain 14 centre-to-limb positions. Those positions are defined as $\mu = \cos \theta$, and include disc positions from the disc centre ($\mu$ =1) to the limb at $\mu$\,=\,0.2. The heliocentric angle, $\theta$, is defined as the angle between the perpendicular line to the solar surface and the line of sight from the observer. Each observation covers a field of view of 32.5\arcsec ($\approx$ 23.600~km) in diameter on the solar disc. The atlas is available online\footnote{\url{https://www.astro.physik.uni-goettingen.de/research/solar-lib/}}.

For this work, we used a modified version of the IAG quiet Sun CLV atlas, as the original IAG CLV atlas contains small gaps where Earth's telluric lines had been removed. To obtain spectra without these gaps, the original data were reprocessed without removing the telluric lines. As the spectra for the respective $\mu$-positions get added, this new version of the data exhibits broadened telluric lines. This artificial broadening occurs because the individual spectra were recorded under varying conditions, causing shifts in the positions of the telluric lines relative to each other.

To make it comparable to the other disc-integrated atlases, this resolved Sun atlas was turned into a disc-integrated atlas following the procedure laid out in \citet{Pietrow2023Nessi}. Specifically, a circular disc was populated with interpolated centre-to-limb variation data, which was limb-darkened following \citet{Neckel1994}. Each cell's profile was shifted according to a differential rotation law by \citet{Horst1986}, after which the disc is integrated into a single full-disc spectrum.

This atlas is uniquely quiet because the observations that make up each $\mu$-bin were taken in areas of the Sun where the Sun was quiet.
We have made both the spatially resolved solar atlas, which includes the telluric lines, and the disc-integrated versions of this atlas available with the publication of this paper.

\subsection{SOLAR-HRS spectra:\ \texorpdfstring{\citet{Meftah2023RemS...15.3560M}}{Meftah et al. 2023}}\label{atlas:meftah}
The SOLAR high-resolution extraterrestrial reference spectra (SOLAR-HRS) were constructed based on existing ground and space observations, encompassing disc-integrated, disc-centre, and intermediate cases. High spectral resolution solar line data from the Quality Assurance of Spectral Ultraviolet Measurements in Europe Fourier Transform Spectrometer (QASUMEFTS) spanning wavelengths from 3000~\AA\ to 3800~\AA\, along with data from the Solar Pseudo-Transmittance Spectra (SPTS) covering 3800~\AA\  to approximately 4400~\AA\, were normalised to the absolute irradiance scale of the SOLAR-ISS reference spectrum. The SOLAR-ISS reference spectrum was developed in 2017 from data obtained from the International Space Station (ISS) \citep{Meftah2018A&A...611A...1M}. It covers a wavelength range from 5~\AA\ to 30000~\AA\ and was used as a lower resolution but very accurate reference spectrum for these normalised spectra.

The SOLAR-HRS spectra were derived by taking the ratio of the high-resolution atmospheric data with the lower-resolution but more accurate ISS data. This process involves five technical steps outlined in Section~4 of \citetalias{Meftah2023RemS...15.3560M}.
Various spectra are generated using this method, each tailored to specific purposes. Technical details of both the SOLAR-HRS and Merged Parallelised Simplified (MPS) ATLAS spectra are summarised in Table~3 of the same paper. The SOLAR-HRS aims to serve as a comprehensive high-resolution solar reference spectrum, representative of solar minimum conditions. The exact dates of the observations are not known or stated in the academic literature, but the data was taken during the solar minimum of 2008 or 2019 (M. Meftah, Personal communication, 17-03-2024). The disc-integrated spectrum has resolutions ranging from 0.01 to 10~\AA\ across the 5 to 44000~\AA\ wavelength range. However, the reported resolution for the 5000-20000~\AA\ wavelength range is between 0.004 and 0.04~\AA. Since the dates of this atlas are not known, no representative disc or spot numbers can be given. The data for this atlas have been made available in digital format \footnote{\url{http://bdap.ipsl.fr/voscat_en/solarspectra.html}}.

\begin{figure*}
    \centering
    \includegraphics[width=0.99\linewidth, trim={0cm 17cm 0.3cm 15cm},clip]{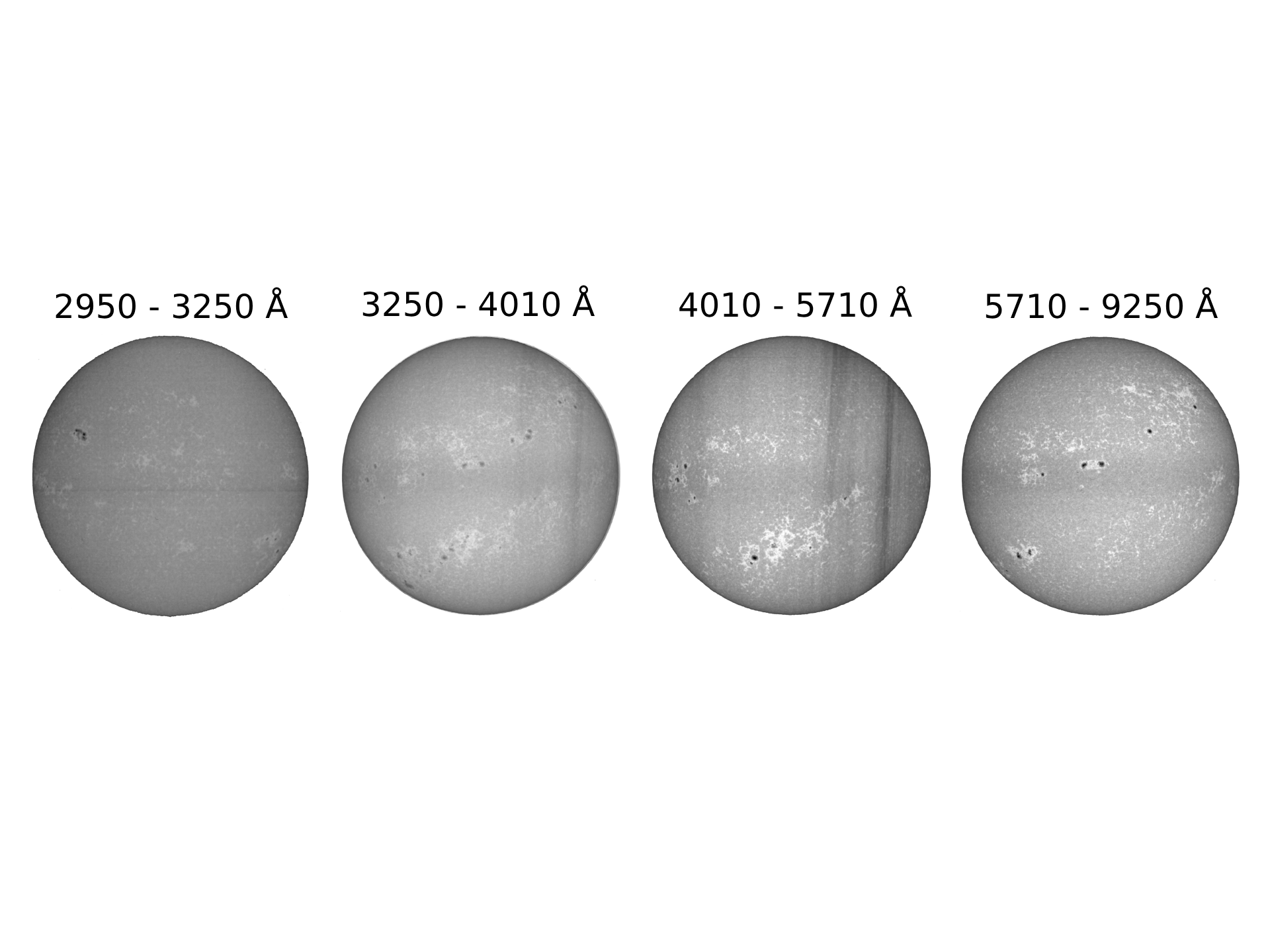}
    \caption{Illustrations of the solar disc during the times of the observations for the Wallace flux atlas, observed by the Meudon spectroheliogram \citep{Malherbe2019, Malherbe23}. The images shown are taken in the Ca~II~K line core, for the times when the atlas observations were taken for each wavelength band, which makes-up the complete Wallace flux atlas. The images contain observational artefacts seen as nearly vertical stripes, most likely caused by clouds.}
    \label{fig:atlascomparewallace}
\end{figure*}

\begin{figure*}
    \centering
    \includegraphics[width=0.99\linewidth, trim={0cm 17cm 0.3cm 15cm},clip]{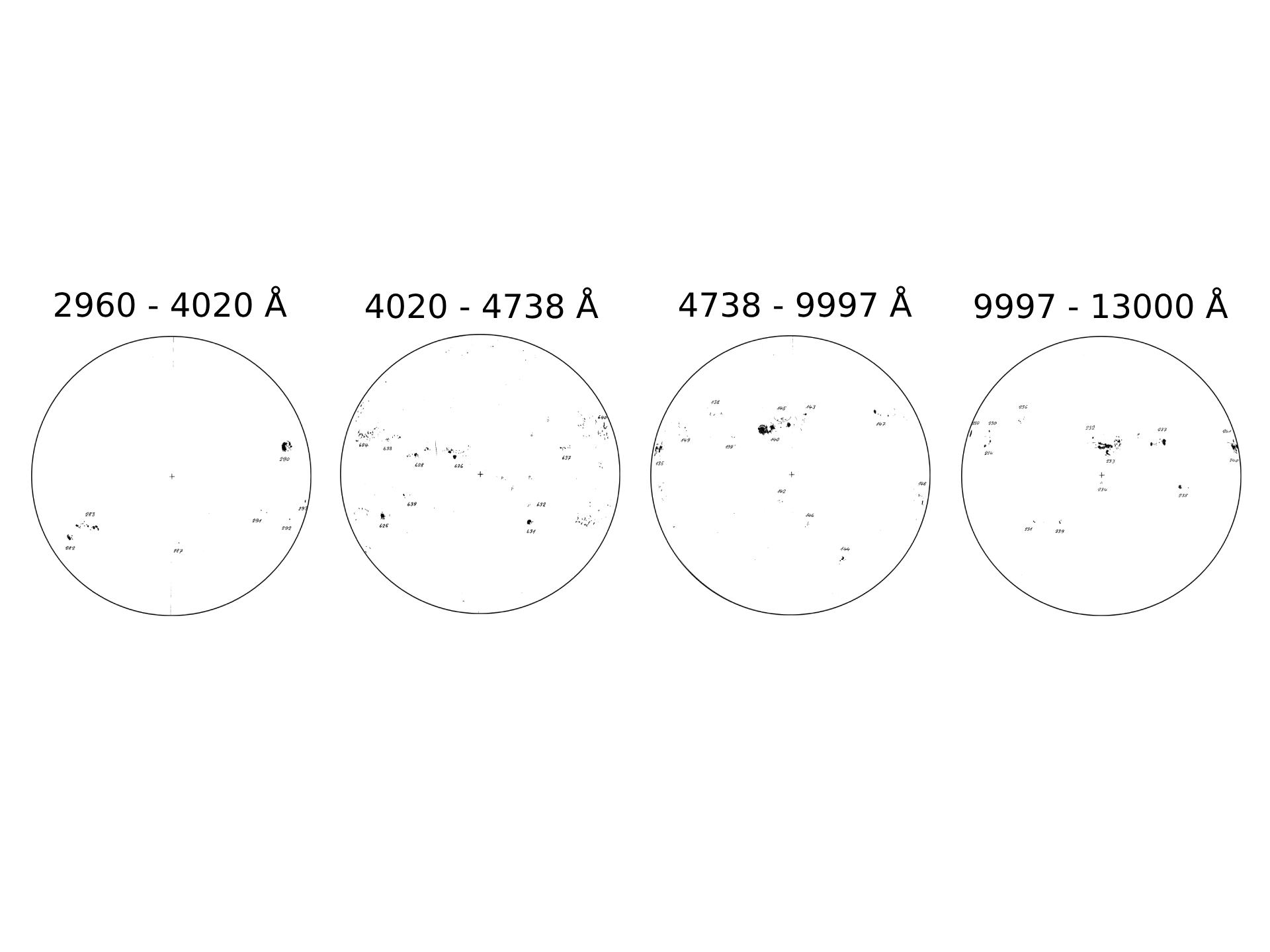}
    \caption{Sunspot drawings in white light, made at the Specola Solare Ticinese\protect\footnotemark during the observations for the Neckel flux atlas. We use the same image to represent the solar disc for several combined wavelength bands that are separated in the atlas. In those cases, the observations were taken within one day of each other and, thus, the disc image is reasonably representative of the solar disc for all those bands.}
    \label{fig:atlascompareneckel}
\end{figure*}
\addtocounter{footnote}{0}
\stepcounter{footnote}
\footnotetext[11]{\url{https://www.specola.ch/en/sunspot-drawings-of-the-specola-solare-ticinese/sunspot-drawings-of-the-specola-solare-ticinese/}}\addtocounter{footnote}{-1}
\subsection{PEPSI Sun-as-a-star: \texorpdfstring{\citet{Strassmeier2018A&A...612A..44S}}{Strassmeier et al. 2018}} \label{atlas:pepsi}
This full-disc atlas spans the wavelength range from 383~nm to 914~nm, divided into six cross-dispersers (CDs) and recorded with a spectral resolution of up to $R \sim 270,000$. The observations were taken using the Solar Disk Integration (SDI) telescope, a 13 mm robotic solar telescope, which feeds the PEPSI spectrograph \citep[see also][for a more detailed description of the spectrograph]{Strassmeier2015} at the Large Binocular Telescope Observatory (LBTO) in Arizona \citep{Hill2012}. The data were collected over three consecutive days, from November 15 to 17, 2016, with up to 112 individual exposures per wavelength setting. The exposures were taken at varying integration times for each CD to maximise the S/N, reaching up to $S/N \sim 8600$ in the red spectral regions. The final deep spectrum was created by combining these exposures for each wavelength range. As stated by the authors, wavelength calibration using Th-Ar exposures and simultaneous Fabry-P\'erot etalon (FPE) combs achieve an absolute accuracy of $10~\mathrm{m\,s^{-1}}$ (root mean square) relative to the more accurate \citetalias{Molaro2013A&A...560A..61M} laser comb atlas (see Section.~\ref{atlas:molaro}), with a daily relative precision of $1.2~\mathrm{m\,s^{-1}}$.

Unlike other atlases designed to reflect specific activity levels, this atlas represents an average solar disc under typical conditions during the declining phase of solar cycle 24, a period of relatively low activity. The author’s own comparisons with \citetalias{Wallace2011ApJS..195....6W} generally reflect this lower activity conditions. The PEPSI atlas, corrected for the solar barycentric motion and gravitational redshift, achieves excellent intensity and wavelength calibration. However, the authors note that for wavelengths shorter than 400 nm, continuum placement becomes challenging due to line blanketing and limited S/N, resulting in slightly larger uncertainties in this range. To mitigate this, the authors used the telluric-corrected \citetalias{Wallace2011ApJS..195....6W} atlas as a continuum guide. 

A representative disc image of this atlas is shown in Fig. \ref{fig:atlascompare}, illustrating the solar conditions during the observations. The data, reduced with the PEPSI-specific processing pipeline, are publicly available from the institute's website \footnote{\url{https://pepsi.aip.de/?page_id=549}}.

\subsection{IAG solar flux atlas:\  \texorpdfstring{\citet{Reiners2016}}{Reiners et al. 2016}} \label{atlas:reiners}
This full-disc atlas consists of two segments, the visual (VIS) part, which ranges from 4050 to 10650~\AA, and the near-infrared (NIR) part, which ranges from 10000 to 23000~\AA. Both are observed using an FTS that is attached to the 50~cm VVT\footnote{\url{https://www.uni-goettingen.de/en/grunddaten/434372.html}} on the roof of the Institut für Astrophysik und Geophysik, Göttingen. The spectrograph is operated in a vacuum and has a spectral resolution of up to \text{R~$\sim$~10$^6$}. The VIS spectrum is composed of 1190 solar observations taken over nine different days between March and June 2014. The NIR spectrum is composed of 1130 solar observations taken over six different days between February and June 2014 \citepalias[see Table 2 of][]{Reiners2016}. We also note that the observations were captured during varying solar activity levels, with more scans taken on days of higher solar activity. The final spectrum is generated by taking the mean value at each wavelength point across the observations with different activity levels (See Fig.~\ref{fig:atlascompare}). The authors utilised the line list from \citet{AllendePrieto1998A&AS..129...41A} to measure the positions of 1252~Fe~I lines, determining the blueshift directly from the \citetalias{Reiners2016}. 

As stated by the authors, this atlas was never meant to function as a quiet Sun atlas, but rather one of the average or even active Sun. This is in contrast to the centre-to-limb (CLV) atlas created by \citet{Ellwarth2023} using the same instrument. This CLV atlas uses data from limited fields of view, but care was taken only to observe the quiet Sun. The representative discs for the two parts of this atlas can be found in the middle two columns of Fig.~\ref{fig:atlascompare}. As stated by the authors, this atlas represents an average across various levels of activity. The data are available on the institute
website\footnote{\url{https://www.astro.physik.uni-goettingen.de/research/flux_atlas/}},
as well as a telluric corrected version created by \citet{Baker2020}.

\begin{table}
    \centering
    \captionsetup{justification=raggedright,singlelinecheck=false}
    \caption{Settings and spot numbers for all six atlases used in this study.} 
    \begin{tabular}{p{1.2cm}p{2.6cm}p{1.5cm}}
        \hline 
        \hline
         Atlas & \phantom{00000}$\lambda$-range [\AA] & Spot Nr \\ 
        \hline
        QS & \phantom{0}3900\: --- \phantom{0}6830 & \phantom{00}0 \\
        \hline
        MA & \phantom{0}3900\: --- \phantom{0}6830 & \phantom{0}67 \\
        \hline
        IAG23 & \phantom{0}4200\: --- \phantom{0}8000 & \phantom{00}0 \\ 
        \hline 
        HRS23 & \phantom{0000}5\: --- 30000 & \phantom{0}- \\
        \hline 
        PEPSI & \phantom{0}3830\: --- \phantom{0}9140 & \phantom{0}27 \\
        \hline 
        IAG16 & \phantom{0}4050\: --- 10650 & 117 \\ 
        & 10000\: --- 23000 & 124 \\
        \hline 
        LFC13 & \phantom{0}4750\: --- \phantom{0}5860 & \phantom{0}24 \\
        \hline
        KPW11 & \phantom{0}2950\: --- \phantom{0}3250 & 153 \\
        & \phantom{0}3250\: --- \phantom{0}4010 & 194 \\
        & \phantom{0}4010\: --- \phantom{0}5710 & 172 \\
        & \phantom{0}5710\: --- \phantom{0}9250 & 197 \\ 
        \hline
        KPN87 & \phantom{0}2960\: --- \phantom{0}4020 & 153 \\
        & \phantom{0}3330\: --- \phantom{0}3783 & 126 \\
        & \phantom{0}4020\: --- \phantom{0}4738 & 159 \\
        & \phantom{0}4738\: --- \phantom{0}5765 & 181 \\
        & \phantom{0}5765\: --- \phantom{0}9997 & 206 \\
        & \phantom{0}9997\: --- 13000 & 258 \\
        \hline
    \end{tabular}
    \label{tab:spot numbers}
    \vspace{1ex}
    \parbox{\textwidth}{\raggedright \vspace{1ex} \textbf{Notes.} \text{$\Delta \lambda$} is the wavelength range of the atlases.\\ \citet{sidc} was used to obtain the spot \\ numbers (Spot Nr). The spot numbers listed are averages per scan \\ per observing day. Links to observation dates are provided in the\\ subsections for each atlas.}
\end{table} 

\subsection{HARPS LFC atlas - \texorpdfstring{\citet{Molaro2013A&A...560A..61M}}{Molaro et al. 2013}} \label{atlas:molaro}
This laser frequency comb (LFC) solar atlas, \citetalias{Molaro2013A&A...560A..61M}, was generated from HARPS observations that were taken on 2010--10--25, at La Silla Observatory, Chile, which, as mentioned in Section~\ref{sec: HARPS-N} precedes the HARPS-N project. This LFC prototype, also called astro-comb, was developed in 2007 by ESO in collaboration with Menlo Systems and the Max-Planck-Institut für Quantenoptik (MPQ) and is thoroughly detailed in \citet{Wilken2010SPIE.7735E..0TW}. 

HARPS spectral observations were made by pointing the spectrograph towards the Moon’s centre with 86\% illumination at a resolving power of \text{\(R = \lambda/\Delta\lambda = 115,000\)}, which is considerably lower than the resolutions achieved by FTS atlases. However, the accuracy of individual LFC lines is significantly higher and are generally free of instrumental effects. 
This solar flux spectra covers a spectral range from 3800 to 6900~\AA. Due to observed flux differences across the spectrum, a good calibration can only be obtained for a reduced wavelength range (i.e. from 4760 to 5850~\AA). The final wavelength-calibrated spectra fell into two separate HARPS detectors; a blue and a red spectra spanning ranges from 4753.68-5304.09~\AA\ and 5337.26-5860.00~\AA,\ respectively. The authors highlight the precision of the LFC spectrograph and assert that the initial line list, adopted from \citet{MolaroMonai2012A&A...544A.125M}, does not require corrections for convective blueshift or gravitational redshift. However, we note that the spectra have been adjusted to account for both the observer's movement relative to the Moon and the Moon's movement relative to the Sun (i.e.  barycentric motion).

In the last column of Fig.~\ref{fig:atlascompare} the average representative disc of this atlas is shown, which could be interpreted to look much less active than those for \citetalias{Reiners2016} or \citetalias{Strassmeier2018A&A...612A..44S}, but it also sports a large active region on the right side which could affect the RV measurement accuracy. However, the \citetalias{Reiners2016} atlas images are averaged over many observations. We can see there are more regions of activity present over the full disc for the \citetalias{Reiners2016} atlas. The \citetalias{Molaro2013A&A...560A..61M} atlas can be downloaded online from the CDS\footnote{\url{https://cdsarc.cds.unistra.fr/ftp/J/A+A/560/A61/}}.

\subsection{Kitt Peak FTS atlas \#1: \texorpdfstring{\citet{neckel1987spectral}}{Neckel et al. 1987}} \label{atlas:neckel}
The McMath Solar Telescope at the National Solar Observatory in Kitt Peak has been used for solar observations since the 1960s. It also utilises an FTS which was described thoroughly by \citet{Brault1978fsoo.conf...33B}. A number of solar flux atlases were published by the institution, the most commonly used and cited of which is the \citet{Kurucz1984sfat.book.....K}. This atlas was the first high spectral-resolution, high S/N atlas produced. Ever since, a multitude of other disc-integrated and disc-centre solar atlases have been produced at Kitt Peak, including the two atlases used in this paper, \citet{neckel1987spectral} and \citet{Wallace2011ApJS..195....6W}, as described in Section~\ref{atlas:wallace}. 

There has been much confusion regarding the \citetalias{neckel1987spectral} atlas, also known as the 'Hamburg FTS atlas' in the scientific literature, where it is often referred to as the atlas of 'Brault and Neckel (1987)' after a possible mistake in the title of \citet{Neckel1999SoPh..184..421N}. However, this publication does not exist, as was noted by \citet{Doerr2016}. The most likely original source is \citetalias{neckel1987spectral}, which is an unpublished note found on the file transfer protocol (FTP) server of the Hamburg Observatory\footnote{\url{ftp://ftp.hs.uni-hamburg.de/pub/outgoing/FTS-Atlas/}}. In this note it is mentioned that the observations used to produce this atlas are identical to the ones used by \citet{Kurucz1984sfat.book.....K}. As such, we use these and Kurucz's notes to briefly describe the data. 

The data for the \citet{Kurucz1984sfat.book.....K} solar flux atlas were obtained in six
days between November 23, 1980, and June 22, 1981. The exact dates for these observations along with the summary of the observational data were published in Table~1 of \citet{Kurucz2006astro.ph..5029K}, which was the telluric corrected version of the 1984-published observations. For each day between 16 and 36 scans of the integrated sunlight were observed and averaged. For the \citetalias{neckel1987spectral} solar spectrum, the FTS wavelengths were calibrated against the Kitt Peak table \citet{Pierce1974kptp.book.....P} for chosen lines, and the intensities against \citet{Neckel1984SoPh...90..205N}. 

Given the age of the data, no full disc observations could be found which were taken on the days of the observations; we have added sunspot drawings for the observation days of this atlas in Fig.\ref{fig:atlascompareneckel}. However, in Table~\ref{tab:spot numbers} we show the spot number for these days\footnote{The calculation used for "spot number" is shown in Section~\ref{sec:Methods}.}. The recorded spot numbers are very high and on average higher than those in the other atlases. We therefore expect this atlas to exhibit higher levels of contamination in comparison. 

\subsection{Kitt Peak FTS atlas \#2: \texorpdfstring{\citet{Wallace2011ApJS..195....6W}}{Wallace et al. 2011}} \label{atlas:wallace}
This Kitt Peak integrated solar flux atlas was stitched together by concatenating two sets of previously taken scans using the FTS at McMath-Pierce Solar Telescope. The first set were taken between November 1980 and June 1981, which were used in the \citet{Kurucz1984sfat.book.....K} atlas, and the second in October 1989. Both sets of observations were made near the peaks of sunspot activity in solar cycles 21 and 22, respectively. To achieve a sufficiently high S/N and spectral resolution \text{($\sim$350,000 and 700,000)}, the full wavelength range was not observed in a single exposure. Instead, the spectrum was divided into six smaller segments using band-pass filters. The details of these segmented observations are listed in Table 1 of \citetalias{Wallace2011ApJS..195....6W}. These observations span the spectral range from 2958 to 9250~\AA.

The Doppler correction for this atlas was determined empirically. Specifically, the line positions of solar Fe~I lines were measured and adjusted to match the frequencies documented in \citet{1994ApJS...94..221N}. As a result, the gravitational red shifts were accounted for and removed from the atlas. The authors state that the wavelength range from 2958 to 5400~\AA\ has been corrected for, and is thus free of, telluric lines. For the remaining wavelength regions, the telluric line corrections were generally determined by comparisons with previously obtained disc-centre spectra. However, it is noted that this process failed in the centres of the stronger lines and hence in the digitally available data, for the very poorly corrected lines, these points are replaced with zeros. 

A representative disc corresponding to the solar conditions used for the four central bins of this atlas are shown in Fig.~\ref{fig:atlascomparewallace}. We note the four bins shown in this figure and in Table~\ref{tab:spot numbers}, compared to the six listed in Table 1 of \citetalias{Wallace2011ApJS..195....6W}. This is because all wavelength bins observed on the same day share the same sunspot number; for clarity, we combined these in the table. The second wavelength bin (3250--4010~\AA) was observed over three days; therefore, the listed sunspot number of 194 is the average across those days. Unlike the more modern atlases, these are taken with the Meudon spectroheliogram \citep{Malherbe2019, Malherbe23}. These images show how this atlas is not only taken during a period of high activity, but it also has different activity levels for its bands\footnote{The data for this atlas can be downloaded from \href{https://nispdata.nso.edu/ftp/pub/Wallace_2011_solar_flux_atlas/?_gl=1*x46g27*_ga*ODI0MzA1NjIuMTcxNzA3MzMwMw..*_ga_TW89ZBTY53*MTcxOTc2MzIxMS45LjEuMTcxOTc2MzU3MC42MC4wLjA}{https://nispdata.nso.edu/ftp/pub/Wallace2011}.}.

\begin{table*}
    \captionsetup{justification=raggedright,singlelinecheck=false}
    \caption{Average formation heights of the investigated spectral lines.}
    \begin{tabular}{p{1.5cm}cccccccccccc}
    \hline
    \hline
        Line & Ca~II~K & Ca~II~H & He~I~D$_{3}$ & \Halpha{}\ & \Hbeta{}\ & NaD$_{2}$ & Mg~I & NaD$_{1}$ & Ca~I~6162 & Fe~I~6301 & Fe~I~6173 & C~I~5380\\
        \hline
        Formation Height~[km] & 2200 & 2080 & 2000 & 1300 & 1200 & 968 & 910 & 888 & 495 & 314 & 100 & 40\\
         \hline
    \end{tabular}
    \label{tab:formation_heights}
    \vspace{1ex}
    \parbox{\textwidth}{\raggedright \vspace{1ex} \textbf{Notes.} \CaIIHK{}: \citep{Bjorgen2018A&A...611A..62B}, \text{He~I~D$_{3}$}: \citep{delaCruz2019A&A...623A..74D}, \Halpha{} \& \Hbeta{}: \citep{Capparelli2017ApJ...850...36C}, \text{NaD$_{1}$ \& NaD$_{2}$}: \citep{Bommier2016A&A...591A..60B}, Mg~I \& Ca~I: \citep{Sampoorna2017ApJ...838...95S}, \text{Fe~I~6301~\AA}: \citep{Grec2010A&A...514A..91G}, \text{Fe~I~6173~\AA}: \citep{Fleck2011SoPh..271...27F}, C~I: \citep{Joshi2011ApJ...734L..18J}.
    It is important to note that the formation heights provided here are intended to provide a general guide for the readers. Activity and structure variations impact the line formation heights, but will not vary so much that a chromospheric line like \Halpha\ to form at a lower height than a photospheric line like Fe~I~6173~\AA\,. These variations typically preserve the height ordering presented.}  
\end{table*}

\section{Methods} \label{sec:Methods}
Table~\ref{tab:spot numbers} lists the settings used for all the atlases investigated in this study along with their respective spot numbers. Relative Sunspot number, also known as the Wolf Number, is the most widely used indirect solar activity proxy as it closely follows the 11 year solar cycle and for which we have historical records that date back to the advent of the telescope almost 400 years ago. It is defined based on the number of sunspots and sunspot groups on the surface of the Sun \citep{Clette2014SSRv..186...35C, Clette2016SoPh..291.2629C}, expressed as

\begin{equation}
\mathrm{R} = k \times (10 \times N\_g + N\_s),
\end{equation}

where k is a scaling coefficient specific to each observer, \text{$N\_g$} is the number of sunspot groups, and \text{$N\_s$} is total number of individual sunspots\footnote{We note that the average sunspot number is not a physical quantity, but we use it here solely as a rough indicator of solar activity.}.

Our QS atlas has been constructed as the median of 162 quietest days during solar minimum and has an average spot number of zero\footnote{See Table \ref{tab:QS_Spot_Numbers} for observation days of this atlas.}. Therefore, we are confident in using our QS atlas as an absolute reference for comparison against the other atlases discussed in the previous section. We used three methods to compare these solar flux atlases. We first computed the equivalent widths (EWs) of ten widely used spectral lines in solar and exoplanetary physics from each atlas, then compared these lines against our QS and MA atlases by defining an activity number for each line. Additionally, we also computed the RVs of these spectral lines. We describe these methods below.

\subsection{Equivalent widths} \label{subsec: ew}
Since most of our atlases were obtained using different instruments with varying wavelength solutions, we applied a wavelength shift to each individual spectral line. To achieve this, we fitted a Gaussian profile to the spectral lines of all atlases (including the QS atlas) and aligned the wavelengths of their line-core minima (i.e. the points of minimum intensity) with that of our QS atlas. The adjustment was necessary because the same wavelength range was used to compute the EW for the same line across all atlases. The wavelength ranges we used were 0.4~\AA\ for C~I~5379.58~\AA, Fe~I~6173.34~\AA, Mg~I~5172.7~\AA, Mg~I~5183.62~\AA, NaD$_{1}$~5895.94~\AA, and NaD$_{2}$~5889.97~\AA, 0.6~\AA\ for Fe~I~6301.51~\AA, 1~\AA\ for Ca~I~6162.18~\AA, 0.9~\AA\ for H$\beta$~4861.34~\AA,\ and 1.2~\AA\ for H$\alpha$~6562.8~\AA, all centred around the core of the lines. These wavelength ranges were selected based on the morphology of the spectral lines. For narrow lines, the range was chosen in regions where the intensity values were approximately equal to 1 (i.e. the continuum intensity). The broader lines, however, required additional considerations to avoid including large portions of the wings. In the case of the \Halpha\ line, we also excluded the telluric line present on the red side of the core in some of the atlases (see Fig.~\ref{fig:halpha_qsm}). Essentially, the EWs of our broad lines were calculated in a manner similar to the bounded equivalent widths previously computed by \citet{Cauzzi2009A&A...503..577C} and \citet{Pietrow23}. Given the differences in resolution and sampling between the atlases, we linearly interpolated all spectral lines to 400 points before calculating the EWs.

\subsection{Activity number} \label{subsec: an}
We  used our two HARPS-N atlases as references to devise a new activity metric. For consistency, the same wavelength ranges, interpolation, and Gaussian-fitted wavelength shift used for EWs (see Section~\ref{subsec: ew}) were applied. The activity number is defined as the difference between the other atlases and our two reference atlases within the specified wavelength ranges for each spectral line. Although our two reference atlases were observed with the same instrument, they were taken at different times and under different conditions, leading to residual instrumental differences. To minimise these differences, we included a normalising factor to account for the discrepancy between the QS and MA atlases. Specifically, for each spectral line, we first divided the linearly interpolated intensity values of the other atlases by the intensities of the QS atlas over the same range. From this ratio, we subtracted the corresponding ratio of the other atlases to the MA atlas. Next, we subtracted the normalising factor, which is defined as one (QS atlas intensities divided by themselves) minus the ratio of the QS atlas intensities to the MA atlas intensities. Finally, the activity number is calculated using an $l1$~normalisation-like technique: the sum of the absolute intensity values is divided by the number of interpolation points, which is set to 400 for all lines. This is expressed as

\begin{equation}
    \mathrm{Activity~number} = \frac{1}{n} \sum \left| \left(\frac{I_{A}}{I_{QS}} - \frac{I_{A}}{I_{MA}}\right) - \left(\frac{I_{QS}}{I_{QS}} - \frac{I_{QS}}{I_{MA}}\right) \right|\times 1000,
\label{eq:activity_number}
\end{equation}
where $I_{A}$, $I_{QS}$, and $I_{MA}$ are the intensities of the other atlases, QS atlas, and the MA atlas, respectively, while $n$ is the number of points interpolated to 400 for all lines. All activity numbers were multiplied by 1000 for readability.  Appendix \ref{appendix:activity_number} gives a more complete introduction to this metric.

Unlike the EW, which is largely independent of spectral resolution, the activity number is highly dependent on resolution. To address this, we convolved the spectral lines from the other atlases using a Gaussian profile. To minimise the differences between the other atlases and our two reference atlases, an optimal convolution sigma was determined iteratively to minimise the activity number, and this value was then applied to each spectral line. We assume Gaussian instrumental profiles for the convolution, which is the same assumption as was made by \citetalias{Reiners2016} for the instrumental profile of their atlas, and largely correct for HARPS-based spectra, which exhibit a slight asymmetry in their instrumental profiles \citep{milakovic2020, Milakovi2023}. The same is assumed for Kitt Peak FTS-based atlases \citep{Brault1985hra..conf....3B} and the SOLAR-HRS spectra (\citetalias{Meftah2023RemS...15.3560M}).

\begin{figure*}
    \centering
    \begin{subfigure}[b]{0.48\linewidth}
        \centering
        \includegraphics[width=\linewidth]{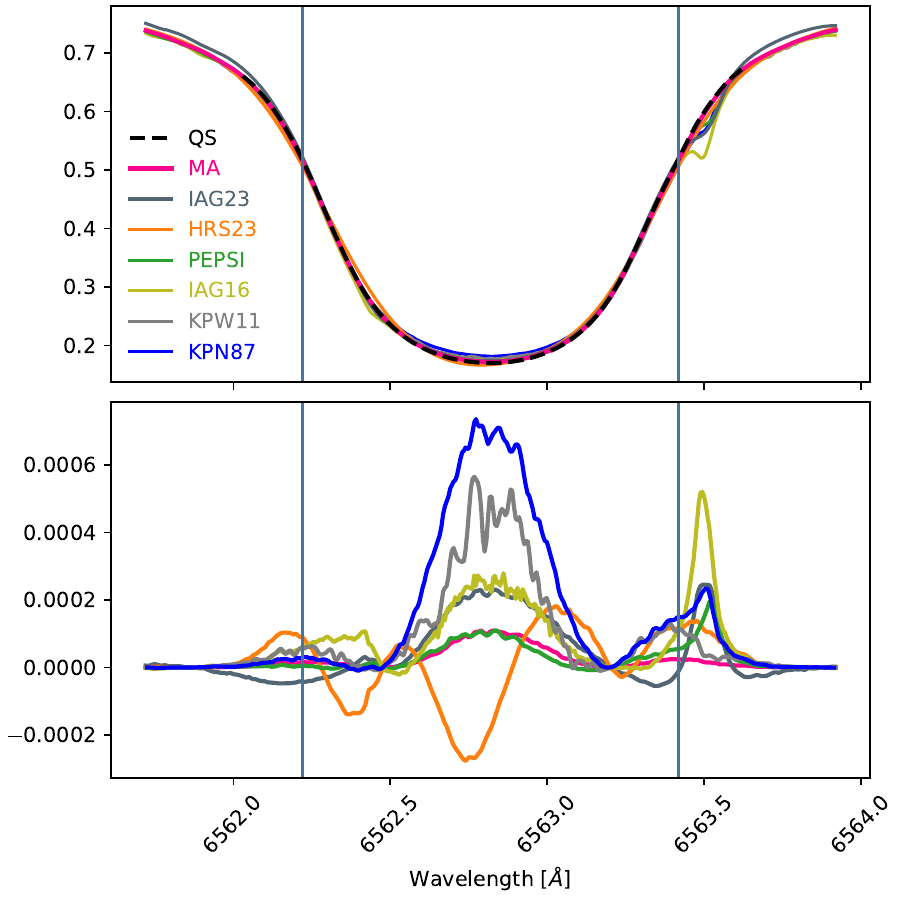}
        \caption{H$\alpha$.}
        \label{fig:halpha_qsm}
    \end{subfigure}
    \hfill
    \begin{subfigure}[b]{0.48\linewidth}
        \centering
        \includegraphics[width=\linewidth]{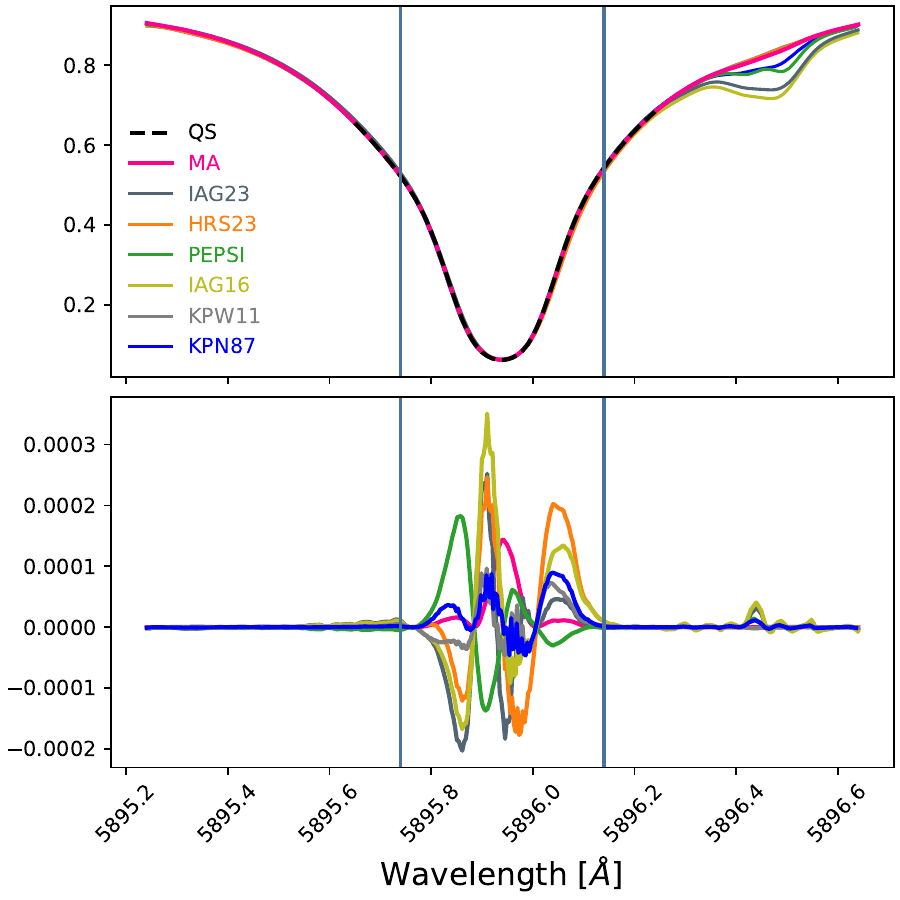}
        \caption{Na\,D1.}
        \label{fig:NaD2_qsm}
    \end{subfigure}
    
    \captionsetup{justification=raggedright,singlelinecheck=false}
    \caption{Illustration of how the activity number method transforms the intensity values of the spectral lines of other atlases. 
    Top panels: the transformed spectral lines after being convolved and interpolated. 
    The two vertical lines indicate the wavelength region used to calculate the activity number. 
    Bottom panels: the remainder of the ratio differences, as described in Section~\ref{subsec: an}. 
    For \citetalias{Meftah2023RemS...15.3560M}, \citetalias{Reiners2016} and \citetalias{Wallace2011ApJS..195....6W} atlases, 
    the wavelengths were converted from vacuum to air using the formula from \citet{Birch1994}.}
    \label{fig:QSMs}
\end{figure*}

For some lines, particularly the two Mg lines, the optimal (i.e. lowest) activity number was achieved by omitting the wavelength shift from the standard procedure, while retaining all other steps. Although \citetalias{Molaro2013A&A...560A..61M} has the same spectral resolution as our HARPS-N atlases, it was the most challenging atlas to match. In addition to omitting the wavelength shift, we also tested removing the Gaussian convolution and applied a small intensity adjustment to its Mg lines. It is important to emphasise that any deviations from the standard procedure were made solely to minimise the activity number. In Figs.~\ref{fig:halpha_qsm} and \ref{fig:NaD2_qsm}, we show two examples of how applying this method transforms the intensity values of the spectral lines.  

\subsection{Radial velocity} \label{radial_velocity}
As an additional method to investigate the varying activity between the different solar atlases, we also determined the RVs of individual spectral lines. To gain a closer understanding of the variance between the atlases, we also examined the bisectors of these spectral lines. The bisectors provide valuable information about the changing solar atmosphere with height. We utilised rest wavelengths from the VALD3 database \citep{K14, Ryabchikova2015} as a reference for this calculation.
\vspace{1ex}The continua of the \citetalias{Meftah2023RemS...15.3560M}, \citetalias{Reiners2016} and \citetalias{Wallace2011ApJS..195....6W} atlases were normalised to unity using the SUPPNet online tool \citep{Suppnet2022A&A...659A.199R} before any of the above methods were computed. 

\begin{table*}
\captionsetup{justification=raggedright,singlelinecheck=false}
\caption{Summary of the EWs [m\AA].}
\begin{tabular}{l|cccccccccc}
\toprule
\hline
 & H$\alpha$ & H$\beta$ & NaD$_{2}$ & NaD$_{1}$ & Mg5173 & Mg5183 & Ca~I~6162 & FeI6301 & FeI6173 & C~I~5380 \\
\midrule
QS & 869.727 & 647.892 & 306.748 & 289.399 & 306.974 & 313.994 & 256.895 & 128.421 & 70.924 & 64.411 \\
IAG23 & 871.091 & 637.254 & 317.717 & 293.237 & 306.684 & 314.560 & 254.763 & 128.452 & 70.881 & 66.082 \\
HRS23 & 875.642 & 647.175 & 310.480 & 295.340 & 308.952 & 318.962 & 256.645 & 128.841 & 71.502 & 63.849 \\
LFC13 & NaN & 643.051 & NaN & NaN & 305.345 & 312.577 & NaN & NaN & NaN & 63.470 \\
PEPSI & 866.067 & 646.339 & 312.663 & 292.301 & 305.272 & 312.575 & 255.045 & 129.046 & 70.201 & 64.102 \\
MA & 869.447 & 648.866 & 306.810 & 289.505 & 306.996 & 314.043 & 256.946 & 128.505 & 70.903 & 64.336 \\
IAG16 & 874.505 & 646.416 & 319.650 & 294.940 & 311.301 & 318.833 & 256.627 & 128.571 & 70.604 & 63.446 \\
KPW11 & 869.871 & NaN & 309.792 & 293.270 & 309.683 & 318.960 & 256.383 & 129.034 & 70.349 & 64.085 \\
KPN87 & 863.721 & 651.776 & 313.400 & 291.927 & 309.283 & 316.935 & 254.782 & 130.414 & 69.902 & 64.818 \\
\bottomrule
\end{tabular}
\label{tab:equivalent_widths}
\vspace{1ex}
\end{table*}

\begin{table*}
\captionsetup{justification=raggedright,singlelinecheck=false}
\caption{Summary of activity numbers}
\begin{tabular}{l|cccccccccc}
\toprule
\hline
 & H$\alpha$ & H$\beta$ & NaD$_{2}$ & NaD$_{1}$ & Mg5173 & Mg5183 & Ca~I~6162 & FeI6301 & FeI6173 & C~I~5380 \\
\midrule
IAG23 & 0.086 & 0.072 & 0.097 & 0.060 & 0.020 & 0.053 & 0.009 & 0.002 & 0.002 & 0.009 \\
HRS23 & 0.101 & 0.079 & 0.032 & 0.085 & 0.069 & 0.068 & 0.003 & 0.002 & 0.001 & 0.003 \\
LFC13 & NaN & 0.083 & NaN & NaN & 0.026 & 0.033 & NaN & NaN & NaN & 0.005 \\
PEPSI & 0.034 & 0.019 & 0.099 & 0.045 & 0.074 & 0.072 & 0.003 & 0.001 & 0.000 & 0.005 \\
MA & 0.035 & 0.018 & 0.018 & 0.024 & 0.007 & 0.009 & 0.001 & 0.000 & 0.001 & 0.000 \\
IAG16 & 0.094 & 0.087 & 0.122 & 0.079 & 0.084 & 0.078 & 0.010 & 0.002 & 0.006 & 0.009 \\
KPW11 & 0.147 & NaN & 0.034 & 0.026 & 0.053 & 0.026 & 0.007 & 0.021 & 0.003 & 0.008 \\
KPN87 & 0.232 & 0.051 & 0.047 & 0.029 & 0.097 & 0.036 & 0.016 & 0.006 & 0.004 & 0.012 \\
\bottomrule
\end{tabular}
\label{tab:activity_numbers}
\vspace{1ex}
\parbox{\textwidth}{\raggedright \vspace{1ex} \textbf{Notes.} The $NaN$ values in Tables~\ref{tab:equivalent_widths} and \ref{tab:activity_numbers}, which primarily appear in the rows corresponding to the \citetalias{Molaro2013A&A...560A..61M} atlas, are due to its limited wavelength coverage. The $NaN$ values in the H$\beta$ columns for \citetalias{Wallace2011ApJS..195....6W} result from missing data.}
\end{table*}

\begin{figure*}[htbp]
    \centering
    \begin{subfigure}[b]{\textwidth}
        \centering
        \includegraphics[width=\textwidth]{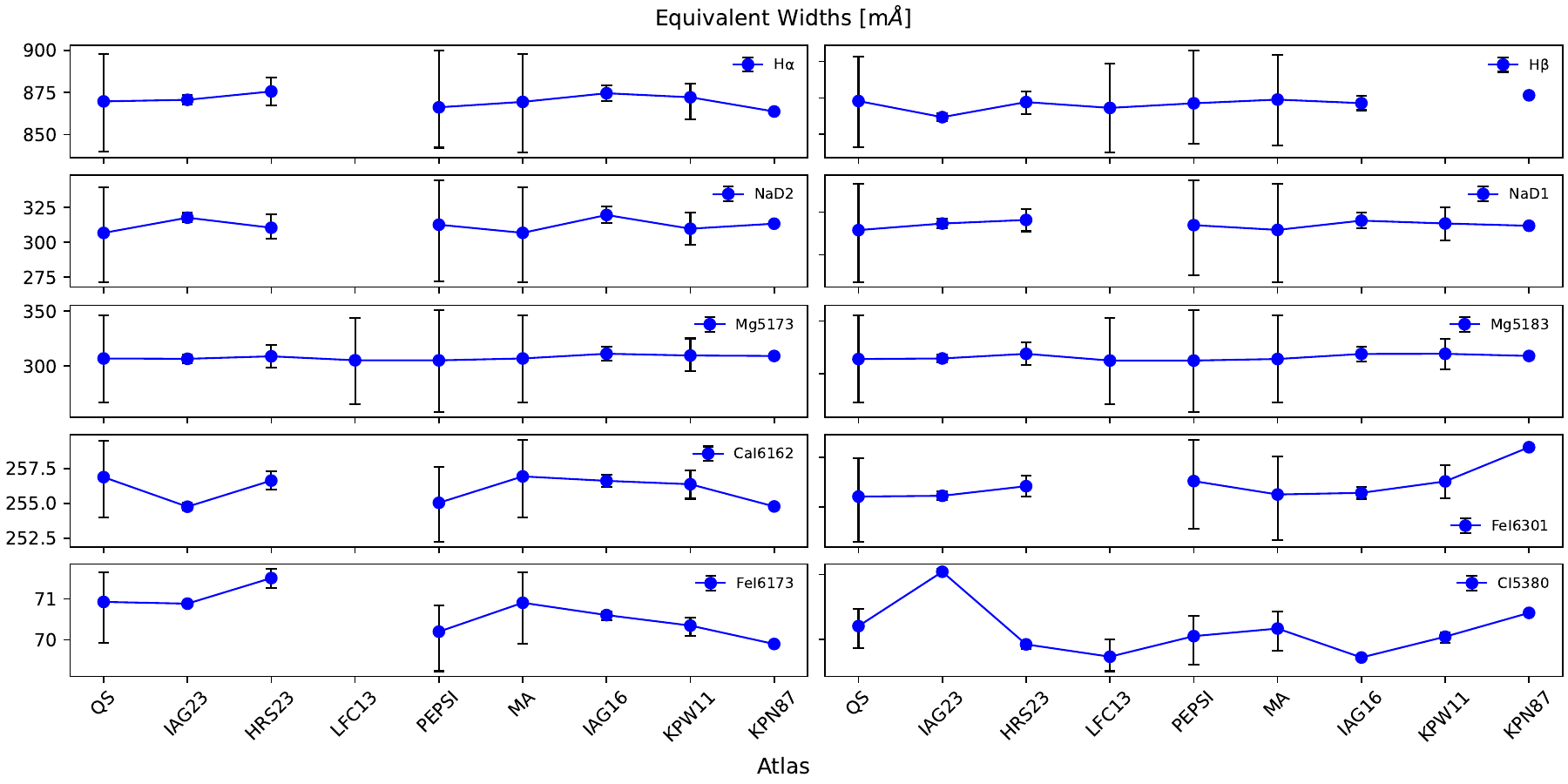}
    \end{subfigure}
    \captionsetup{justification=raggedright,singlelinecheck=false}
    \caption{EWs and the associated uncertainties for all atlases.} 
    \label{fig:All_EWs}
    \hfill
    \begin{subfigure}[t]{\textwidth}
        \centering
        \includegraphics[width=\textwidth]{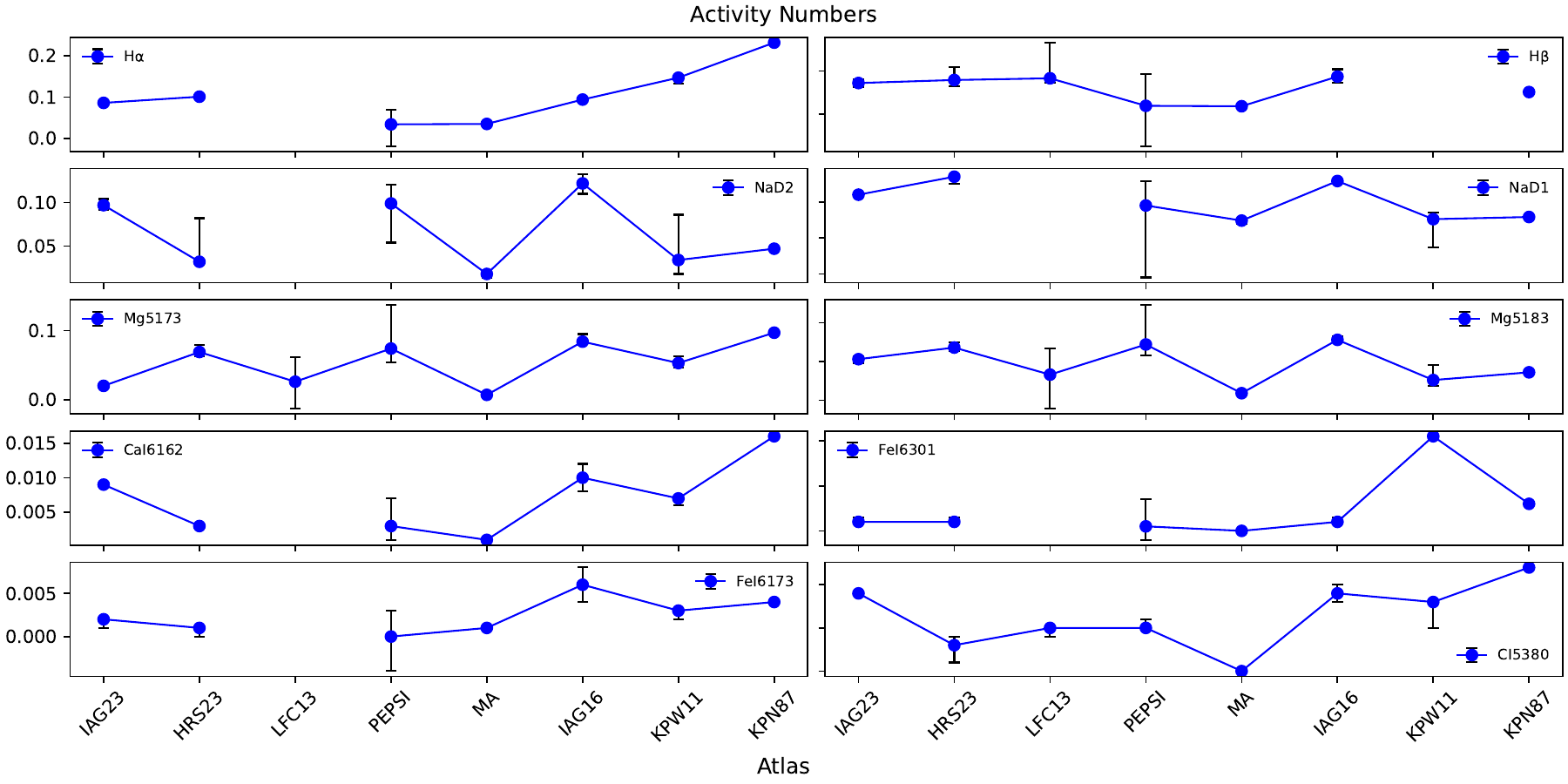}
    \end{subfigure} 
    \caption{Activity numbers along with associated uncertainties for all atlases.} 
    \label{fig:All_ANs}
\end{figure*}

\begin{figure*}[htbp]
    \centering
    \begin{minipage}[t]{0.48\textwidth}
        \centering
        \includegraphics[width=\textwidth]{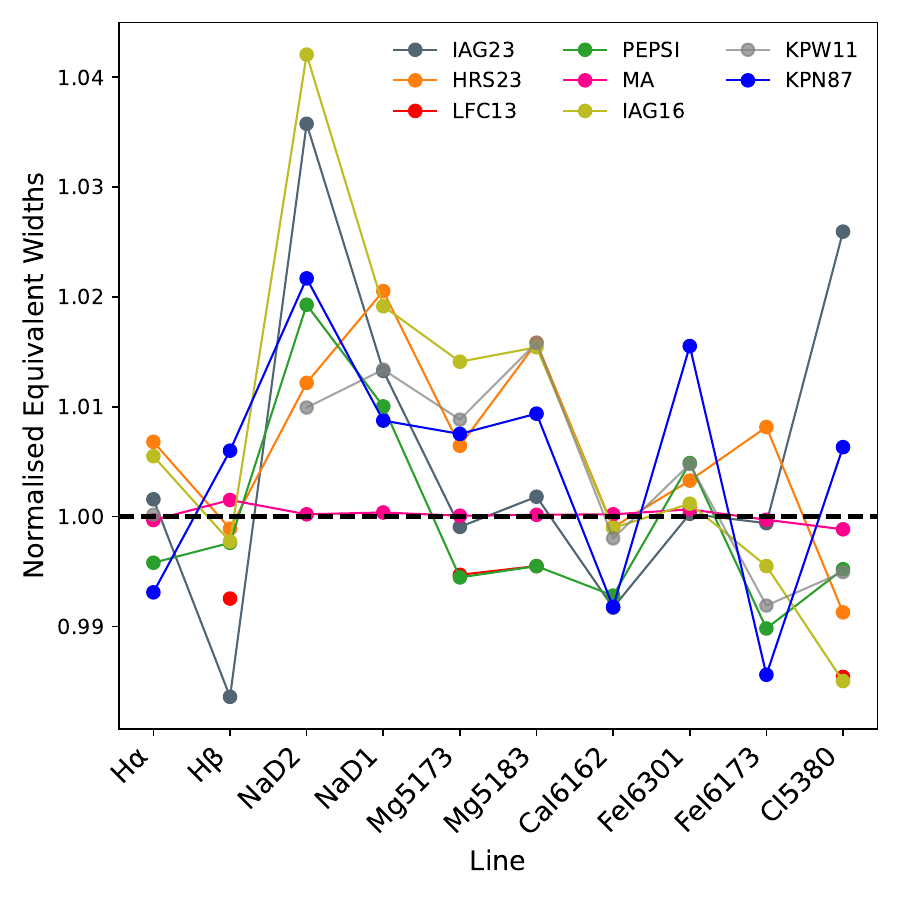} 
        \captionsetup{justification=raggedright,singlelinecheck=false}
        \caption{Normalised EWs per atlas plotted against spectral lines ordered by average core formation height. Error bars have been omitted for visual clarity.}
        \label{fig:EWs_per_atlas}
    \end{minipage}
    \hfill
    \begin{minipage}[t]{0.48\textwidth}
        \centering
        \includegraphics[width=\textwidth]{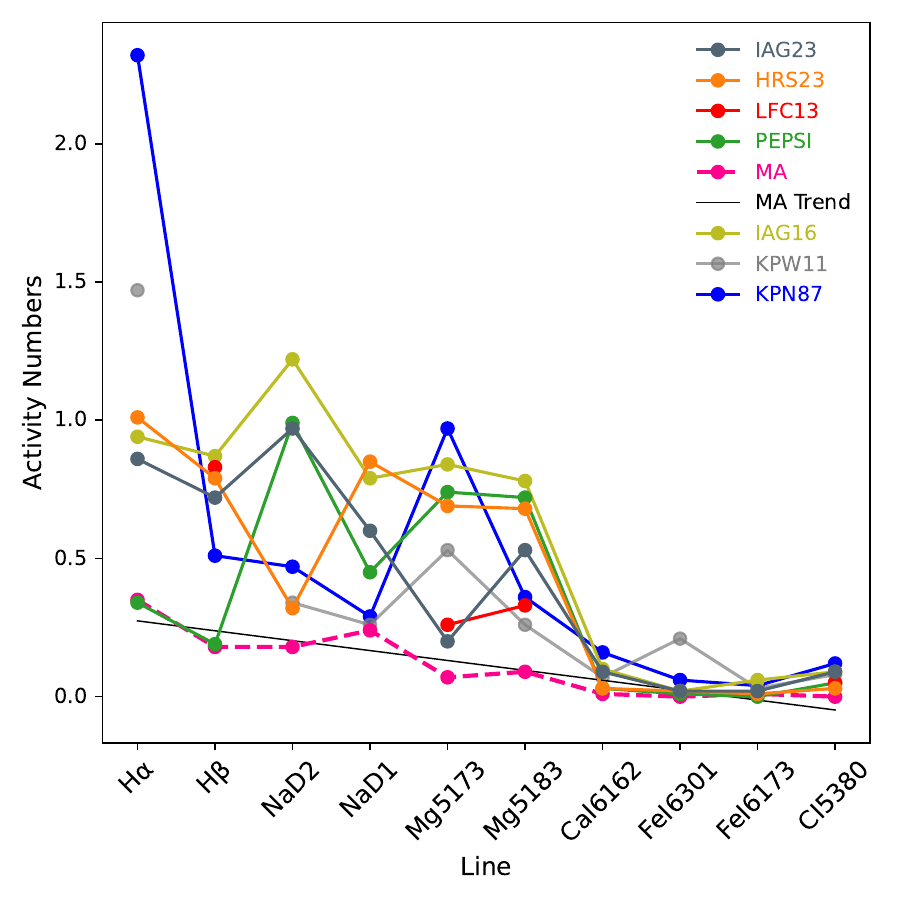}
        \captionsetup{justification=raggedright,singlelinecheck=false}
        \caption{Activity numbers per atlas plotted against spectral lines ordered by average core formation height. Activity numbers of our MA atlas as a function of core formation height have a linear trend (solid black line). Again, error bars have been omitted for visual clarity.}
        \label{fig:AN_per_atlas}
    \end{minipage}
\end{figure*}

\section{Results} \label{sec: results}
The EWs and activity numbers obtained, as described in the previous section, are presented Tables~\ref{tab:equivalent_widths} and \ref{tab:activity_numbers}. We use our QS atlas as a benchmark reference and true representative of the quiet Sun conditions. As  Fig.~\ref{fig:atlascompare} shows, this is the most quiet atlas with an average spot number of zero (see Tables~\ref{tab:spot numbers} and \ref{tab:QS_Spot_Numbers}). Throughout the paper, the tables are arranged in order of increasing sunspot number (see Table~\ref{tab:spot numbers}), with the atlas representing the lowest activity level at the top and the highest activity level at the bottom. The spectral lines are also ordered by average core formation height (see Table~\ref{tab:formation_heights}). 

\subsection{Solar activity} \label{subsec:4.1}
Measuring EWs is a widely used activity metric assumed to be independent of spectral resolution. Conversely, our new activity number is computed relative to two reference atlases and thus is highly dependent on spectral resolution and instrumental differences (see Appendix.~\ref{appendix:activity_number} for a detailed inspection of this newly devised metric and its interpretation). Our MA atlas, as indicated by its relative spot number (Table.~\ref{tab:spot numbers}) and activity during observation time (Fig.~\ref{fig:atlassolarcycle}), should exhibit higher levels of activity than \citetalias{Ellwarth2023}, \citetalias{Meftah2023RemS...15.3560M} and \citetalias{Molaro2013A&A...560A..61M}. In Figs.~\ref{fig:All_EWs} and \ref{fig:All_ANs}, we present the EWs and activity numbers, respectively, plotted separately for each spectral line across all atlases. To estimate uncertainties in the EWs and activity number measurements, we varied the integration range by an amount corresponding to three spectral sampling elements ('pixels'). Specifically, the mean wavelength sampling within the selected wavelength range around each spectral line was measured, and a shift equal to three times this local sampling step was applied. This resulted in two additional measurements with integration ranges reduced and expanded by this amount, respectively. The spread between these estimates and the central measurement was adopted as an uncertainty measure. Our interpolation of the spectra onto a common grid, i.e.\ 400 points, is also a potential source of error; however, this uncertainty is quite difficult to quantify.
\subsubsection{Correlation with spot number} \label{subsec:4.1.1}
Before undertaking this study, we anticipated observing a rise in activity as a function of spot number. For this reason, the atlases throughout this manuscript are ordered by spot number, which serves as an indirect measure of the level of activity on the Sun's surface. However, as it is clear from the plots in Figs.~\ref{fig:All_EWs} and \ref{fig:All_ANs} and contrary to our initial expectations, we found no measurable trend or correlation between the activity levels in solar flux atlases and their corresponding spot numbers for most spectral lines.  

To investigate this further, we computed the Spearman correlation coefficients ($\rho$) and their associated p-values for the EWs and activity numbers. For the EWs, only the Fe~I~6301.51~\AA\ line shows a statistically significant positive correlation, with $\rho = 0.8095$ and a p-value of 0.0149. In contrast, the Fe~I~6173.34~\AA\ line exhibits a statistically significant negative correlation, with $\rho = -0.7143$ and a p-value of 0.0465. For the activity numbers, the only line with a statistically significant trend is the \Halpha\ line, which has $\rho = 0.8286$ and a p-value of 0.0416. Interestingly, the two iron lines have the second and third highest $\rho$-values for activity numbers, but their p-values are not statistically significant.

Sunspots are strongly correlated with the S-index and the Mg~II index \citep[see, for example,][]{Maldonado2019A&A...627A.118M}. However, other magnetic activity features, such as plages and faculae, have been shown to induce larger variations in spectral lines \citep{Meunier2010A&A...512A..39M, Carlsson2019ARA&A..57..189C, Pietrow2020A&A...644A..43P}. Nevertheless, a stronger correlation between activity and spot number in atlases would still be expected; however, in our results, this relationship is likely obscured by the much more dominant instrumental effects (see Section~\ref{subsec:4.1.3}). Although it is noteworthy to mention that the pseudo-emission features in the relative spectra produced from HARPS-N data by \citet{Thompson2020} exhibited a strong correlation with facular regions but no correlation with sunspots.
 
\subsubsection{Correlation with core formation height} \label{subsec:4.1.2}
It is well known that broad lines, whose cores typically form in the chromosphere and whose wing formations span much wider spectral regions, are significantly more sensitive to activity than narrow lines, whose cores primarily form in the photosphere \citep{Livingston2007ApJ...657.1137L, GomesdaSilva2011A&A...534A..30G, Maldonado2019A&A...627A.118M}. Line-by-line comparisons of EWs must be treated with utmost caution because stronger broad lines form significantly deeper cores and have much wider wings than narrow lines, which naturally results in higher EWs. Our activity number, on the other hand, has the advantage of being independent of the core formation depth because it is calculated relative to a reference atlas, the QS atlas. Therefore, the activity number provides a more reliable metric for comparing the relative sensitivity of spectral lines to solar activity. This distinction is seen in Figs.~\ref{fig:EWs_per_atlas} and \ref{fig:AN_per_atlas} where the activity number captures the much higher variability amplitudes of the broad lines, a feature that is largely absent in the EWs.

\begin{figure*}
    \centering
    \begin{subfigure}[b]{0.48\linewidth}
        \centering
        \includegraphics[width=\textwidth]{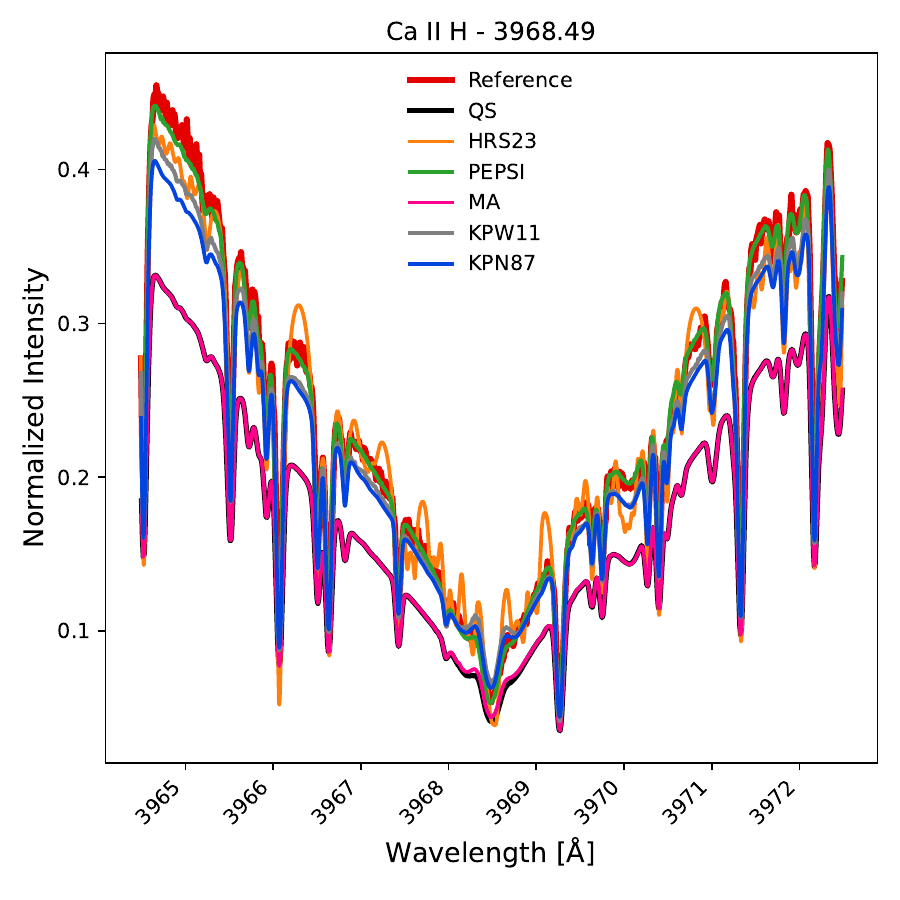}
        \caption{}
        \label{fig:CaH}
    \end{subfigure}
    \hfill
    \begin{subfigure}[b]{0.48\linewidth}
        \centering
        \includegraphics[width=\textwidth]{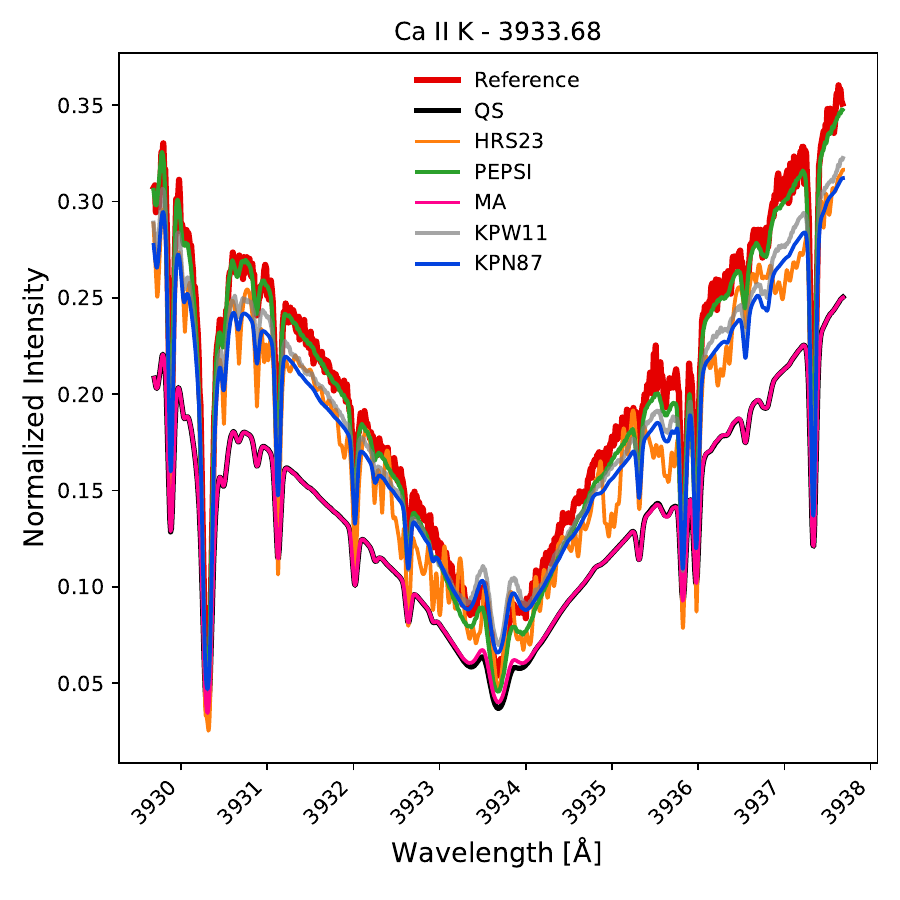}
        \caption{}
        \label{fig:CaK}
    \end{subfigure}
    \captionsetup{justification=raggedright,singlelinecheck=false} 
    \caption{ \CaIIHK\ lines of all the atlases that cover this region overplotted together. The plots clearly show the intensity difference between our QS and MA atlases and the other atlases.}
    \label{fig:All_CaLines}
\end{figure*}

In stark contrast, the behaviour seen in Fig.~\ref{fig:AN_per_atlas} shows that, from the left-hand side (i.e. from \Halpha\ to the Mg~I~5183 line), activity numbers exhibit sporadic and high-amplitude variability. However, all atlases converge at the Ca~I~6162~\AA\ line, where both the activity and its variability remain consistently small in comparison. With the exception of \citetalias{Molaro2013A&A...560A..61M} due to its limited wavelength coverage, the correlation with average core formation height is very robust for both metrics, as indicated by the Spearman correlation coefficients ($\rho$) and their associated p-values. However, the activity numbers of our MA atlas stand out, as they follow a clear linear trend (see the solid black line in Fig.~\ref{fig:AN_per_atlas}). This result is particularly noteworthy because there are no instrumental differences between the MA and QS atlases. While differing observational conditions may introduce some discrepancies, these are relatively minor compared to the instrumental differences present in the other atlases. Therefore, a significant portion of the differences between the MA and QS atlases can be attributed to solar activity. As the spectral line plots in Figs.~\ref{fig:All_CaLines}~\&~\ref{fig:All_figures}d suggest, this linear trend likely extends further to the upper chromospheric helium line, He~I~D$_{3}$~5875.62~\AA, and the \CaIIHK\ lines, whose cores typically form at higher heights than \Halpha\, see Table.~\ref{tab:formation_heights} (We will discuss the \CaIIHK\ lines in detail in Section~\ref{sec:CaLines}).  

It should be noted that, despite using a wider spectral window for the Ca~I~6162~\AA\ line than for most broad lines, its EW and activity number remain much smaller. This highlights its lack of sensitivity to activity compared to the significantly higher sensitivity of broad lines.

\subsubsection{Impact of instrumental profiles} \label{subsec:4.1.3}
The EWs and activity numbers of our MA atlas provide valuable insight into the challenges of quantifying solar activity. We deliberately chose to compare the EWs of spectral lines across the different atlases under investigation, based on the presumption that this metric is independent of spectral resolution and instrumental profiles. However, our results reveal an unexpected outcome: the EWs of the MA atlas are nearly identical to those of the QS atlas. Specifically, the average fractional difference between the MA and QS atlases is only about 0.05\%, indicating an almost perfect match across all lines. By comparison, the other atlases show substantially larger deviations from both the MA and QS atlases, with average fractional differences for all lines typically in the range of 0.7–1.2\%. This range is even larger for the broad lines alone, reaching 0.6-1. 6\%, representing an order of magnitude increase in variability. The EWs of the four atlases with the lowest spectral resolution (QS, MA, \citetalias{Molaro2013A&A...560A..61M}, and \citetalias{Strassmeier2018A&A...612A..44S}) show the largest uncertainties (see Fig.~\ref{fig:All_EWs}), meaning that small changes in the integration range can have a disproportionate impact on these atlases due to their larger wavelength steps. This indicates that the chosen wavelength range has a significant effect on the stability of the EWs in lower-resolution atlases. Similarly (and as expected), our activity numbers also suggest a strong dependence on spectral resolution and instrumental profiles. The activity numbers of the MA atlas, especially for the broad lines, are substantially lower than those of most other atlases.  

In addition to highlighting the difficulty of quantifying solar activity, our findings underscore the considerable impact of instrumental profiles on the morphology of spectral lines. In theory, by applying a wavelength shift to the other atlases, as described in Section~\ref{subsec: ew}, we attempted to remove the offsets introduced by the different wavelength scales and calibrations inherent to the various atlases. However, other differences in line shapes, invisible to the naked eye, remain due to the respective observing instruments and cannot be easily eliminated. \citetalias{Reiners2016} compared the wavelength scales of their FTS atlas with two Kitt Peak FTS atlases, \citet{Kurucz1984sfat.book.....K} and \citetalias{Wallace2011ApJS..195....6W}, as well as comparing the two Kitt Peak atlases with each other. Despite these two Kitt Peak atlases being observed with the same instrument and setup, the authors found offsets as large as $600~\mathrm{m\,s^{-1}}$ in the blue region of the spectrum and up to $200~\mathrm{m\,s^{-1}}$ in the red regions due to differences in calibration methods. Therefore, we expect errors in the calibration techniques used in \citet{DiazBaso2021zndo...5608441D} as well as \citet{Lofdahl2021} who used the disc centre atlases of \citet{Neckel1999SoPh..184..421N} and \citet{Brault1978fsoo.conf...33B} as reference for their calibrations.  

The spectral resolution of \citetalias{Molaro2013A&A...560A..61M} is the same as that of our two HARPS-N atlases, and it was observed under quieter solar conditions compared to the MA atlas. As mentioned earlier, we applied both wavelength and intensity shifts to its lines to align them with our two atlases. Nevertheless, its EWs show significant deviations, and its activity numbers are much larger than those of the MA atlas. While the solar activity in each line is contained in the values listed in Table~\ref{tab:activity_numbers}, it is evident that the largest portion of these values arises from instrumental differences. This raises the question: how can we disentangle the exact contributions of solar activity from the instrumental imprints in these values?  

The wavelength scale of \citetalias{Molaro2013A&A...560A..61M} is presumably more accurate. Recent attempts have been made to improve the accuracy of spectral calibrations using LFC technology \citep{Jallageas2019, Reiners2024}. Even so, such methods cannot eliminate the unique and nuanced imprints of different observing instruments on spectral lines, which our findings demonstrate to be the greatest challenge in comparing spectral lines across different atlases.

\begin{figure*}[htbp]
    \centering
    \includegraphics[width=0.98\textwidth]{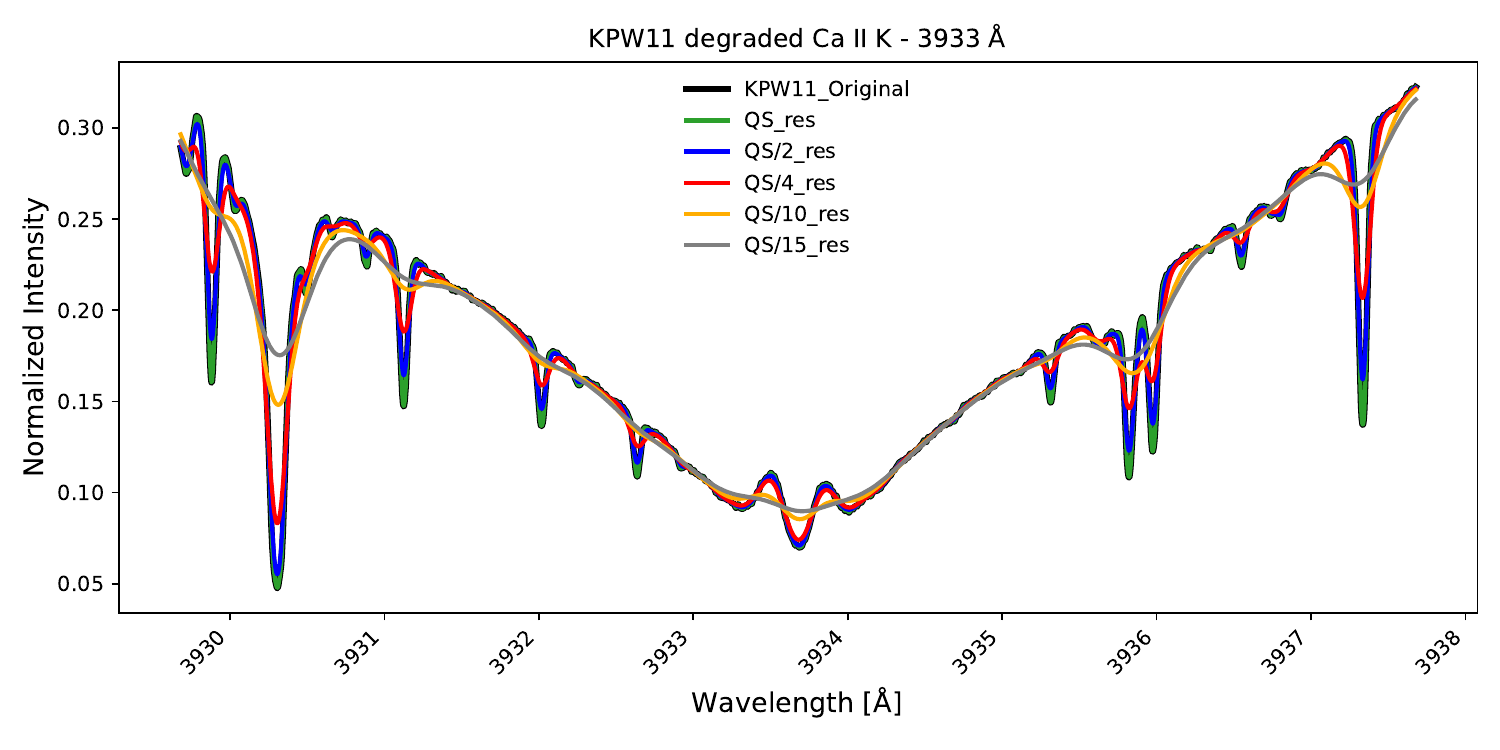} 
    \captionsetup{justification=raggedright,singlelinecheck=false}
    \caption{\citetalias{Wallace2011ApJS..195....6W} atlas degraded to the resolution of our QS atlas and lower around the Ca~II~H~3933.68~\AA\ line. The resolution was lowered by performing a Gaussian convolution while keeping the spectral sampling intact.}
    \label{fig:KPW11_degraded}
\end{figure*}

\begin{figure*}
    \includegraphics[width=0.99\linewidth]{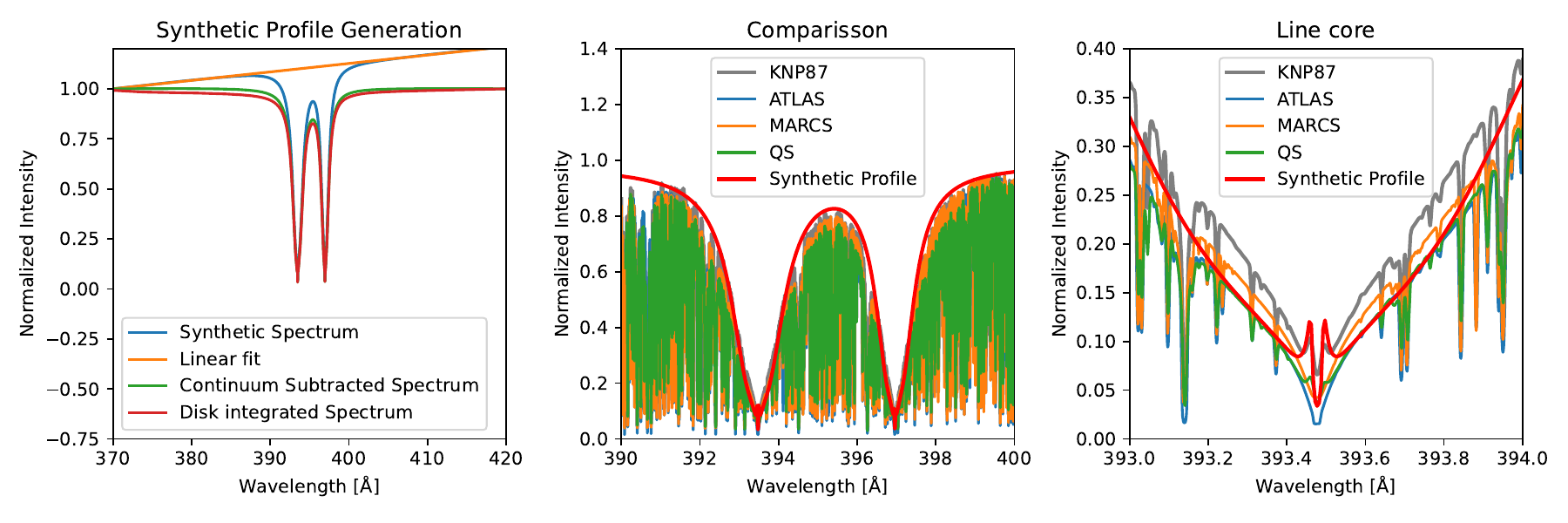}
    \caption{Comparison of the \CaIIK\ lines of our QS atlas, \citetalias{neckel1987spectral}, MARCS, and ATLAS spectra with a quiet Sun FAL-C atmosphere synthesised profile.}
    \label{fig:SyntheticCaLines}
\end{figure*}

\subsection{\texorpdfstring{\CaIIHK{}}{CaIIHK}} \label{sec:CaLines}
The \CaIIHK{} doublet lines which occur in the violet part of the visible spectrum at 3968.49 and 3933.68~\AA,\ respectively, are two of the most studied chromospheric lines. Their extended wings span the photosphere \citep{Sheminova2012SoPh..280...83S, Cretignier24} and their cores form at similar heights in the mid-chromosphere, with the K line forming at a slightly higher height due to its greater opacity \citep{Bjorgen2018A&A...611A..62B}. Ground-based observations in the violet end of the spectrum, below $\sim$~4000~\AA, are hampered by two primary aspects. First, the Earth's atmosphere, which absorbs a significant portion of the light in this region, reducing the intensity of the signals that reach ground-based telescopes \citep{Ermolli013ACP....13.3945E}. Second, low sensitivity of optical detectors, which makes it harder to detect and accurately measure the wavelengths in these regions while also observing other lines \citep[see Section 2 of][for a review]{Chatzistergos2024JSWSC..14....9C}. 

In Fig.~\ref{fig:All_CaLines}, we overplot the \CaIIHK{} lines of all the atlases that cover this region of the spectrum. As made evident here, there is a significant intensity difference between the \CaIIHK{} lines of our QS and MA atlases compared to the other atlases. This highly unexpected intensity difference and continuum placement relative to our HARPS-N atlases is why we have omitted these lines from our EW and activity number calculations. However, this issue certainly warrants further investigation. Naturally, one might ascribe this difference to either the levels of activity in other atlases or the lower resolution of our HARPS-N atlases. Clearly, if the level of activity significantly impacts the continuum placement, it would be noticeable in the MA atlas. For example, the Ca~II~K line of the MA atlas closely matches the QS atlas, with a mean intensity difference of just 0.57\%. The highest difference, 3.44\%, occurs at the far red wing of the line, though this is dwarfed by a 41.86\% difference with \citetalias{Wallace2011ApJS..195....6W}. Again, it is evident that activity level alone cannot explain such a large difference. 

Generally, instruments with relatively high resolving powers, \text{> 100,000}, resolve the spectrum well enough to avoid any issues on the continuum level \citep{Bouchy2001}. Some issues may occur only for very low resolutions \text{$\sim$~1000}, which is not the case with HARPS-N. To test this, we degraded the \citetalias{Wallace2011ApJS..195....6W} atlas, assuming a Gaussian profile, to the resolution of our QS atlas and much lower (see Fig.\ref{fig:KPW11_degraded}). Clearly, lower resolutions of the order of the resolving power of HARPS-N have no discernable impact on the continuum levels or even the line shape. We also refer the readers to Figure.~1 of \citetalias{Reiners2016} who degraded their atlas to the resolution of the \citetalias{Molaro2013A&A...560A..61M} atlas, which has the same resolution as our atlases, and found no impact on the continuum level in their atlas.    

This leaves the normalisation methods used in constructing the various atlases as the only other avenue for investigation. Our two HARPS-N atlases are in total disagreement with all other atlases, which generally show very strong agreement with each other. However, since our atlas is normalised to a 3D local thermal equilibrium (LTE) synthetic solar spectrum from the POLLUX database \citep{Palacios2010} (see Section. \ref{sec: HARPS-N}), the question arises regarding how the others are normalised. 

\begin{figure}[htbp]
    \centering
    \includegraphics[width=0.99\linewidth]{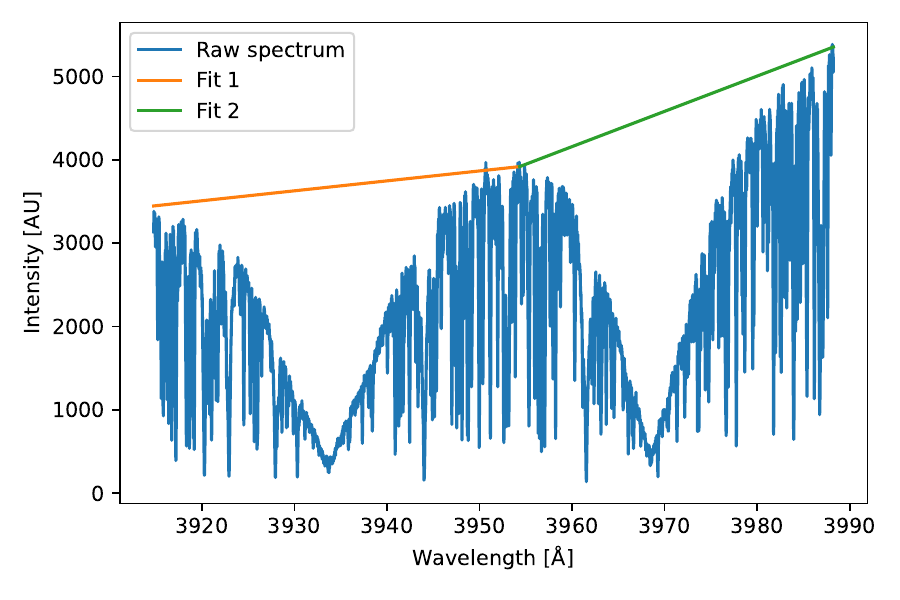}
    \captionsetup{justification=raggedright,singlelinecheck=false}
    \caption{Illustration of how we linearly connected two random high points on either side of the \CaIIHK{} lines to obtain Reference in Fig.~\ref{fig:All_CaLines}.}
    \label{fig:ca_normalised}
\end{figure}

The \citet{Kurucz1984sfat.book.....K} atlas is the most cited atlas produced at Kitt Peak. The physical book that came with the spectrum includes a note related to Fig. 3, namely, that the pseudo-continuum is far from being the real continuum for scans 1-7 (wavelength ranging from \text{$\sim$~2960} to 4700~\AA). The authors also explain that for these wavelength ranges a pseudo-continuum was fitted to the highest point out of every 100 points, and then each scan was normalised to this pseudo-continuum. A slightly different normalisation technique was used for \citetalias{neckel1987spectral}'s atlas which is described in \citet{Neckel1984SoPh...90..205N}. However, in a similar fashion when describing the polygonal tracks below 4420~\AA\ the authors state that they should be seen as tracks connecting the incidental highest maxima set  according to the personal choice of the authors, rather than as qualified approximations of the 'real 'solar continuum'. The original normalisation methods introduced in the Kitt Peak atlases, especially the one adopted in \citet{Kurucz1984sfat.book.....K}, appear to have been widely accepted and propagated over the years, subsequently establishing the \CaIIHK{} continuum levels as a reference for newer atlases. This explains the strong correlation between the lines of all other atlases, and the difference with our independent normalisation.

As stated above, our atlases were constructed using the POLLUX database with no direct input from us. Thus, our normalisation is independent of the other atlases, and POLLUX has the advantage of providing synthetic spectra further into the blue spectrum than the \CaIIHK{} lines, and as such fits the continuum beyond them more easily \citep{Palacios2010}. To further test our \CaIIHK{} lines we compare our lines with synthetic spectra which do not rely on a continuum normalisation assumption.
Figure.~\ref{fig:SyntheticCaLines} shows an image of the \CaIIHK{} lines synthesised at disc centre using the quiet Sun FAL-C atmosphere \citep{Fontenla1993} with the 1D non-LTE (NLTE) partial redistribution (PRD) Lightweaver \citep{Osborne2021} code for a 5 level Ca~II atom. Then it was run through the NESSI code \citep{Pietrow2023Nessi} to create a full-disc spectrum using the limb darkening parameters from \citet{Neckel1994}. Although this approximation is not perfect (e.g. not synthesised in 3D, although these effects have been shown to be negligible \citep{Bjorgen2017}, and using only continuum limb darkening parameters which have been shown to be different in the line core \citep{Pietrow23}), it does show a strong resemblance with our data rather than the \citetalias{neckel1987spectral} atlas profile. 
This continuum of our synthetic line is flat because besides calcium and hydrogen, no other lines were considered in this synthesis. This makes it trivial to fit the continuum as a linear relation suffices. 
This line is depicted in red in all three panels of Fig.~\ref{fig:SyntheticCaLines}. In addition, we have our atlas in green, the 3D LTE MARCS spectrum in orange, and the 1D LTE Atlas spectrum in blue. From this it is clear that while the \citetalias{neckel1987spectral} spectra fit better in the far wings, there is a large discrepancy between this spectrum and that of the others, including our synthetic spectrum.

In order to complete our investigation, we normalised the original data of our QS atlas (see Fig.~\ref{fig:ca_normalised}) by linearly connecting two randomly selected high points on either side of the \CaIIHK{} lines (Reference in Fig.~\ref{fig:All_CaLines}). This simple procedure produces a line that closely resembles that of \citetalias{Strassmeier2018A&A...612A..44S} and is comparable to the Kitt Peak atlases. However, it is important to emphasise that this normalisation technique lacks any physical basis and, as stated in \citet{Neckel1984SoPh...90..205N}, is a 'personal choice'. The complexity of normalising this region of the spectrum is clearly evident, and continued efforts are needed to improve upon the currently available techniques.

\subsection{Radial velocities of single spectral lines} \label{rv_section}
\begin{figure*}[htbp]
    \centering
    \begin{minipage}[t]{0.49\textwidth} 
        \centering
        \includegraphics[width=\textwidth]{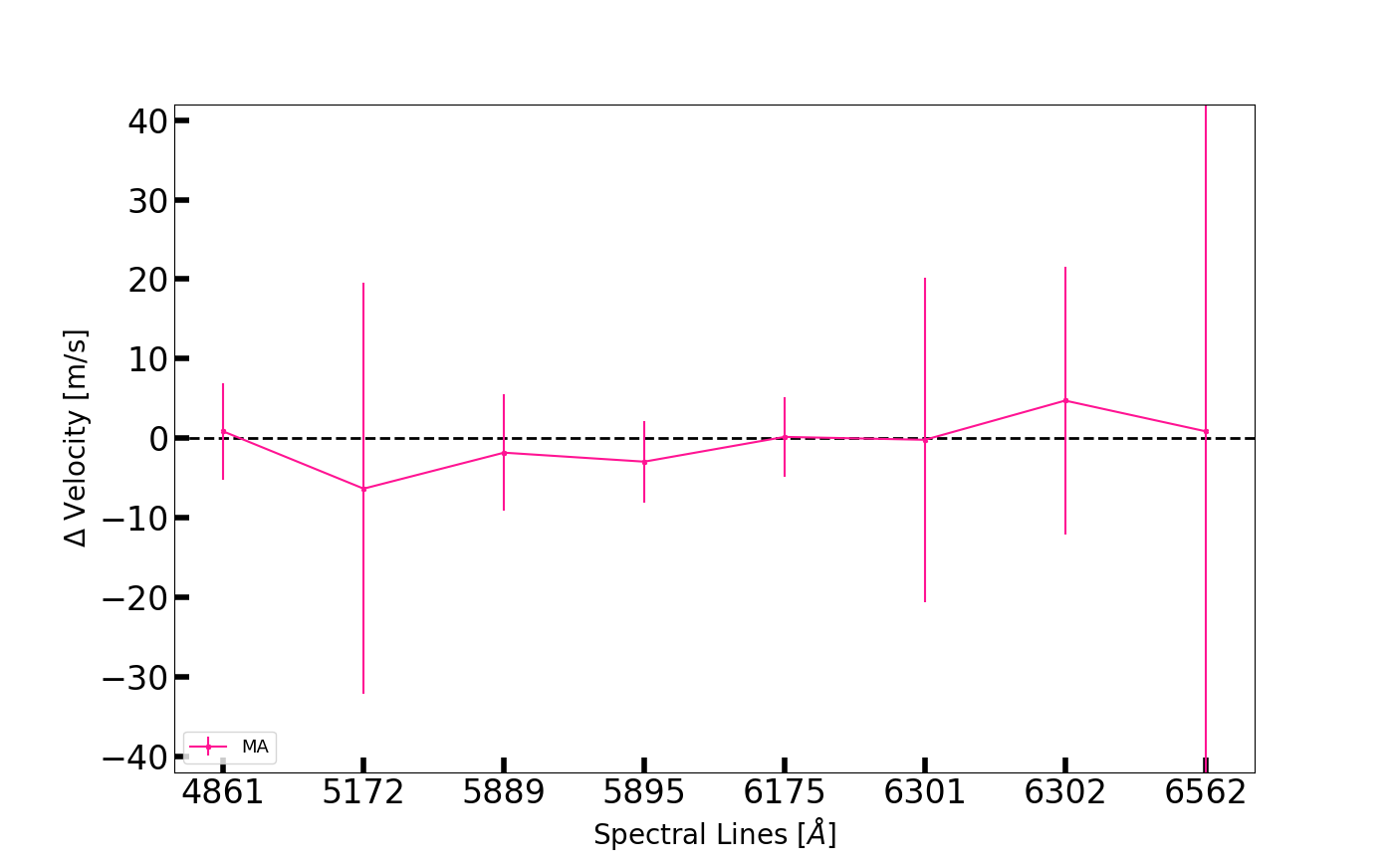}
        \captionsetup{justification=raggedright,singlelinecheck=false}
        \caption{RV deviation of individual spectral lines between the QS and MA solar atlases. The error bars indicate the uncertainties obtained from the line-profile fitting procedure.}
        \label{fig:diff_MA}
    \end{minipage}
    \hfill 
    \begin{minipage}[t]{0.49\textwidth} 
        \centering
        \includegraphics[width=\textwidth]{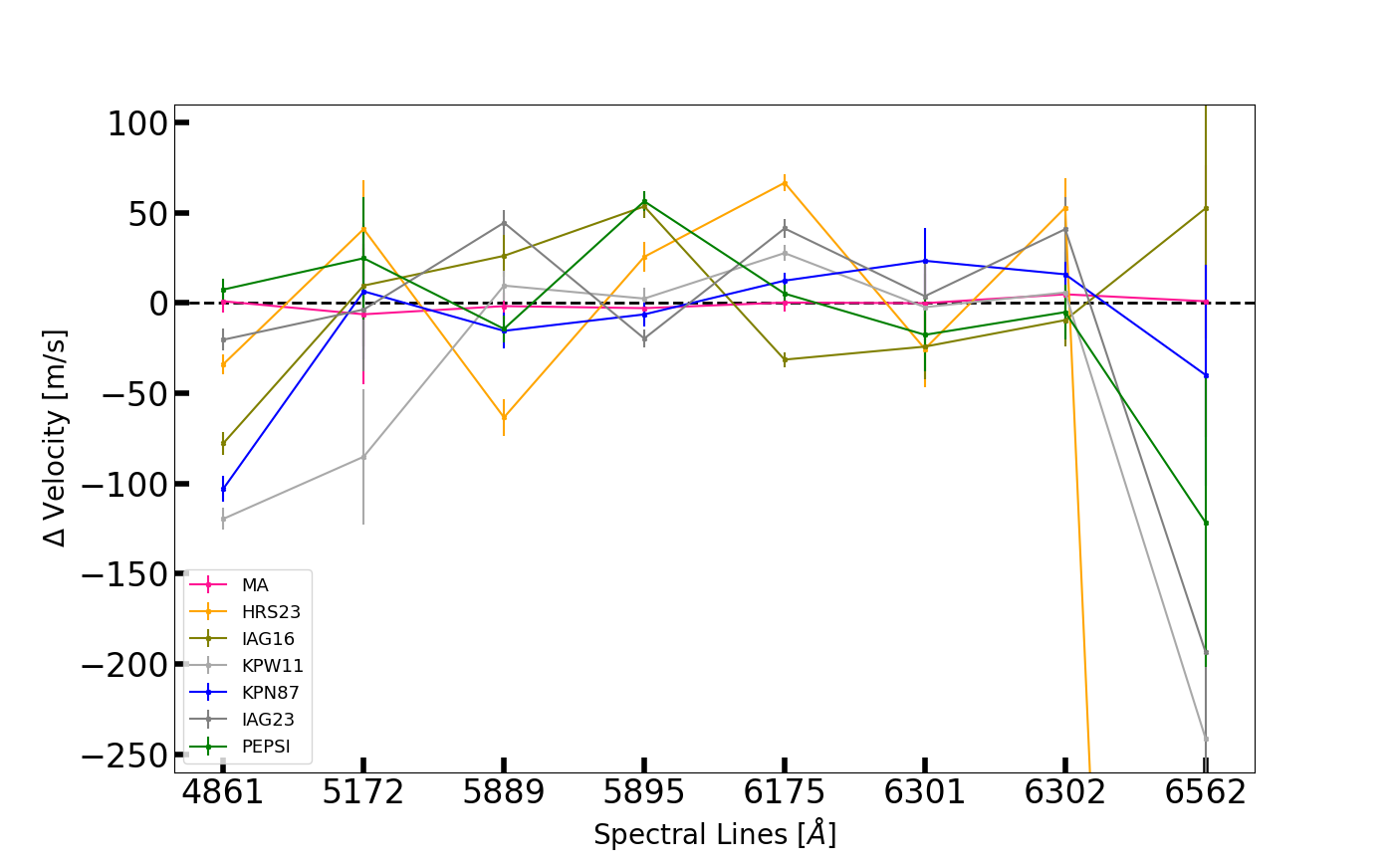}
        \captionsetup{justification=raggedright,singlelinecheck=false}
        \caption{RV deviation of individual spectral lines between the atlases, with the QS atlas serving as the reference. The error bars display the uncertainties obtained from the line-profile fitting procedure. The \citetalias{Meftah2023RemS...15.3560M} RV for the H$\alpha$ line is outside the visible range, as it is an outlier with a deviation in magnitude of 2.}
        \label{fig:diff_all}
    \end{minipage}
\end{figure*}

As the aim of this work was to investigate the variances between the different atlases arising from observations taken during periods of varying solar activity, we also provide a brief overview of the radial velocities (RVs) of individual spectral lines. This analysis allows us to examine in greater detail how these solar atlases differ. We calculated the RVs for eight spectral lines (H$\beta$~4861~\AA, Mg~I~5172~\AA, Na~I~D~5889~\AA, Na~I~D~5889~\AA, Fe~I~6175~\AA, Fe~I~6301~\AA, Fe~I~6302~\AA, and H$\alpha$~6562~\AA), using the rest wavelengths provided by the VALD3 database as a reference.
Figure~\ref{fig:diff_MA} shows that the RVs measured in the MA atlas differ only slightly from those of the QS atlas. This is expected because solar activity typically causes shifts in the range of 10\,m\,s$^{-1}$. However, when comparing the RVs of the other atlases to the QS atlas, the scatter is noticeably larger (see Fig.\,\ref{fig:diff_all}).
Since determining the wavelength solution with high accuracy is challenging, we subtracted the average difference between the QS RVs and the respective atlases. 

This simplification is acceptable here because the goal is not a detailed analysis of the RVs for these individual lines, but rather to obtain a general impression of the scale of variations involved. The measured velocities of all the analysed lines vary in a range of hundreds of m\,s$^{-1}$, highlighting the inherent difficulty in making detailed, direct comparisons between different solar atlases. The variations in the wavelength solutions and line shapes predominantly arise from instrumental and observing conditions. These variances make it more challenging to disentangle solar activity from instrumental effects. Furthermore, some spectral lines exhibit sufficiently large fitting uncertainties, making it challenging to discern whether these uncertainties are due to instrumental effects, inherent limitations in fitting procedures, or intrinsic imprecision of the lines themselves. Given these substantial variations in individual spectral lines, a more comprehensive analysis involving a significantly larger number of spectral lines would be necessary to achieve conclusive 
results. However, such an investigation would go beyond the scope of this paper.

\begin{figure}
    \centering
    \includegraphics[width=\linewidth]{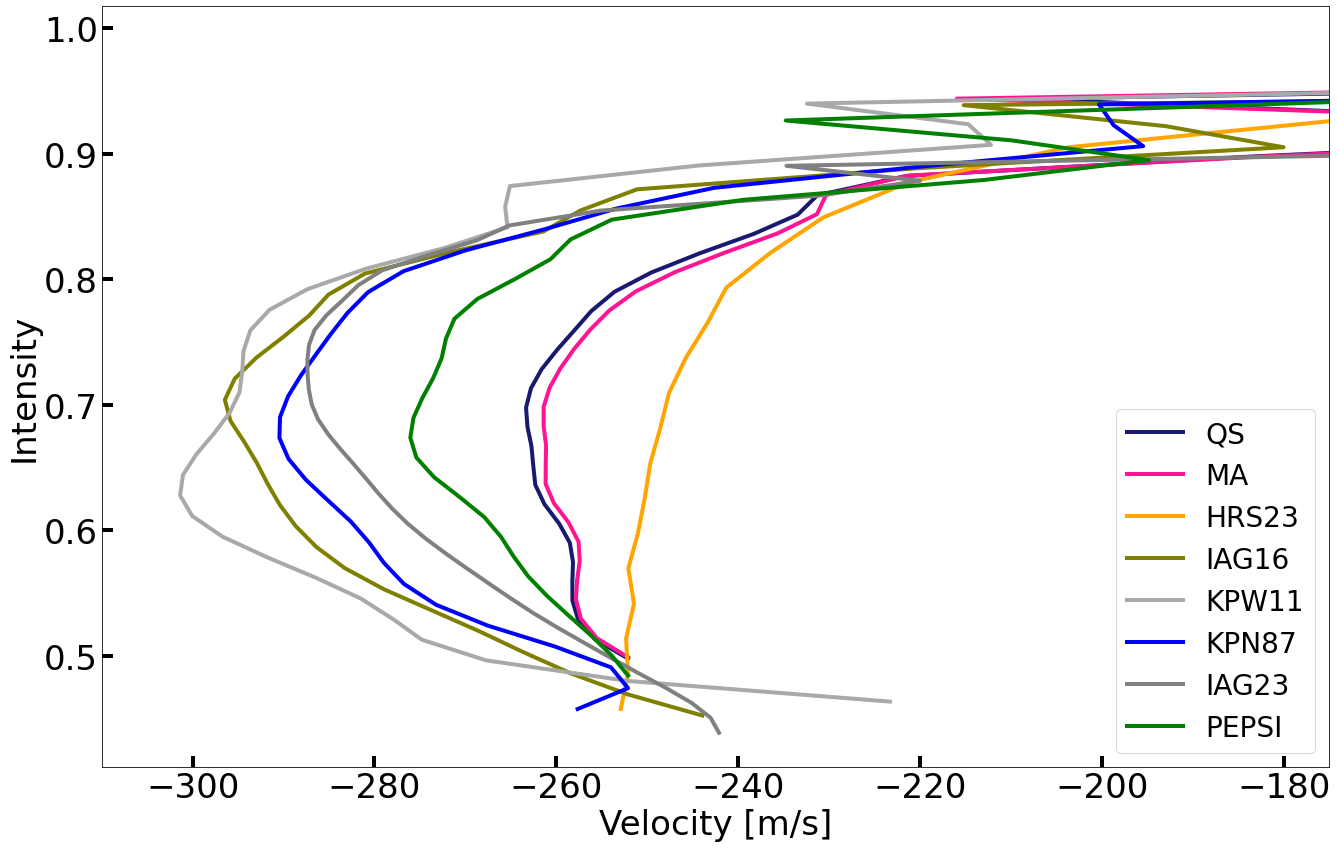}
    \caption{Bisectors of the Fe~I 6173~\AA\ line, shifted towards the RV of the QS atlas. Error bars not included for visual clarity.}
    \label{fig:bis}
\end{figure}

\subsubsection{Bisector investigation of the  Fe~I~6175~\AA\ lines}
To investigate the behaviour of lines in more detail, bisectors can be used. They help to reveal the true asymmetry of lines. The analysis of the bisector of single lines is intended for illustrative purposes only, as individual lines are hardly representative. We carried out the bisector analysis for all the 8 lines used to calculate the RVs in Sec. \ref{rv_section}. However, for illustration, we chose the well-known photospheric Fe~I line at 6173~\AA\,, shown in Fig.~\ref{fig:bis}. This line has a Landé-factor of g$_{eff} = 2.5$ and is therefore highly Zeeman-sensitive. Consequently, we expect to observe changes in this line under increased solar activity. Again, we shifted the velocities towards the QS reference velocity since we are only interested in the general shape. Comparing QS to MA shows a stronger blueshifted bisector (about 10 m\,s$^{-1}$) for the QS, which is expected, since solar activity is mitigating the general blueshift (see e.g. \cite{2016A&A...587A.103L}). 
It is also noticeable that the QS and MA atlases exhibit a less pronounced C-shape in their bisectors compared to the lines from the other atlases (except \citetalias{Meftah2023RemS...15.3560M}), which can be attributed to their lower spectral resolution. We analysed the bisectors of all eight lines from Fig.~\ref{fig:diff_all} and, as expected, the spectral resolution is one of the strongest factors influencing the bisector shape. To further investigate this effect, we degraded the resolution of all eight lines in all atlases to match that of HARPS-N and examined the resulting behaviour. However, this adjustment did not provide new insights as intrinsic instrumental effects appear to be so significant that simply matching the resolution does not make the atlases directly comparable. Again, a more extensive analysis involving hundreds of lines would be needed to address this issue in detail. In this work, we like to emphasise that analysing only a few individual lines does not yield a conclusive result.\\
What we aim to highlight here is that the bisectors of the \citetalias{Meftah2023RemS...15.3560M} atlas show significant discrepancies. For all the bisectors of the \citetalias{Meftah2023RemS...15.3560M} we found an unexpectedly much less pronounced C-shape. Some of the spectral lines of this atlas that we have investigated exhibit blue asymmetries, see Fig.~\ref{fig:HalphaComparison} for example. Observations show dynamic filaments and early phases of flares can cause blue asymmetry in the \Halpha{} line core \citep{Huang2014A&A...566A.148H, Kuridze2015ApJ...813..125K}, which could explain the blue asymmetry in the core of the \Halpha{} line of \citetalias{Meftah2023RemS...15.3560M} seen in Fig.~\ref{fig:HalphaComparison}. However, we note that \citet{Tei2018PASJ...70..100T} only found the Mg~II h and k lines to exhibit this blue asymmetry in a C-class flare. Unfortunately, due to the limited information available on the observation time of this atlas, we are unable to provide a conclusive explanation for this behaviour.

\section{Summary and conclusion} \label{sec:Conclusion}
In this study, we introduce our new HARPS-N QS atlas as a benchmark reference for atlases in the visible range of the spectrum and investigated solar activity using multiple metrics. Our primary aim is to highlight the degree of activity-induced contamination that generally impacts other atlases. As a starting point, we used the relative spot numbers of the atlases to guide our analysis, arranging the atlases in ascending order of spot number, with the expectation that this metric would display a strong correlation with other activity indicators. However, contrary to our expectations, we found no statistically significant correlation between spot numbers and other activity metrics based on their Spearman correlation coefficients.

\begin{figure}[htbp]
    \includegraphics[width=0.99\linewidth]{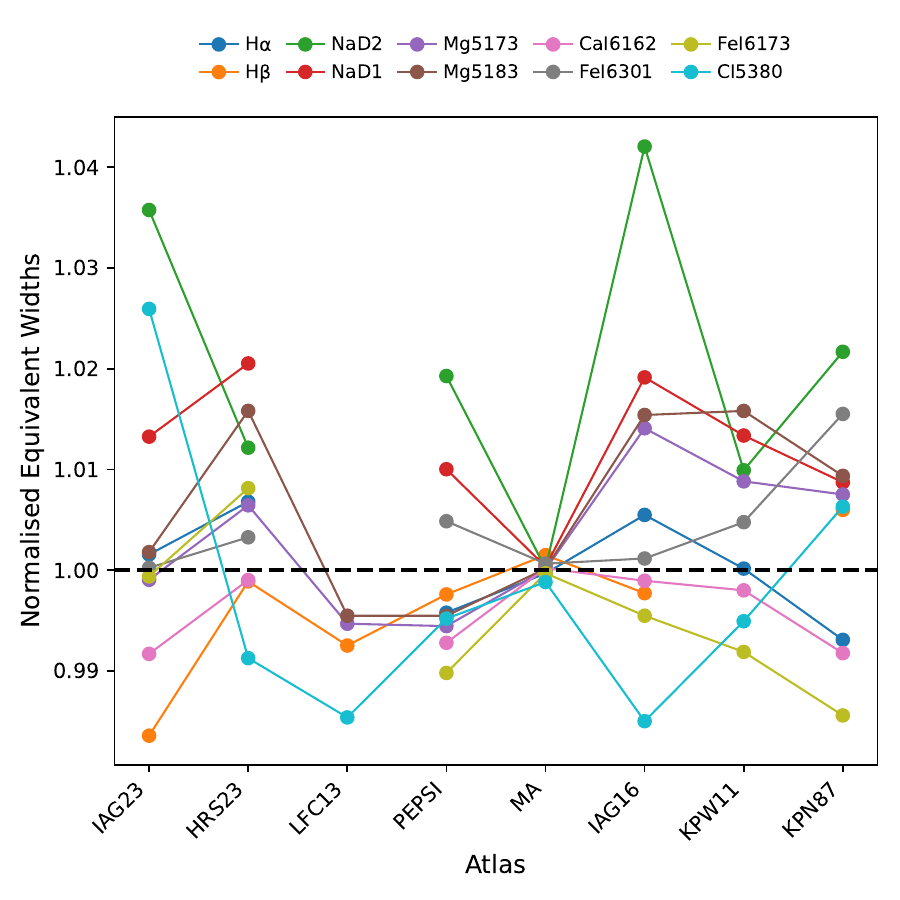}
    \caption{All the EWs listed in Table.~\ref{tab:equivalent_widths} normalised to the EWs of the QS atlas. Error bars have also been omitted here for visual clarity.}
    \label{fig:normalised_ews}
\end{figure}

\begin{figure*}
    \centering
    \begin{subfigure}[b]{0.45\linewidth}
        \centering
        \includegraphics[width=\textwidth, height=0.8\textwidth, keepaspectratio]{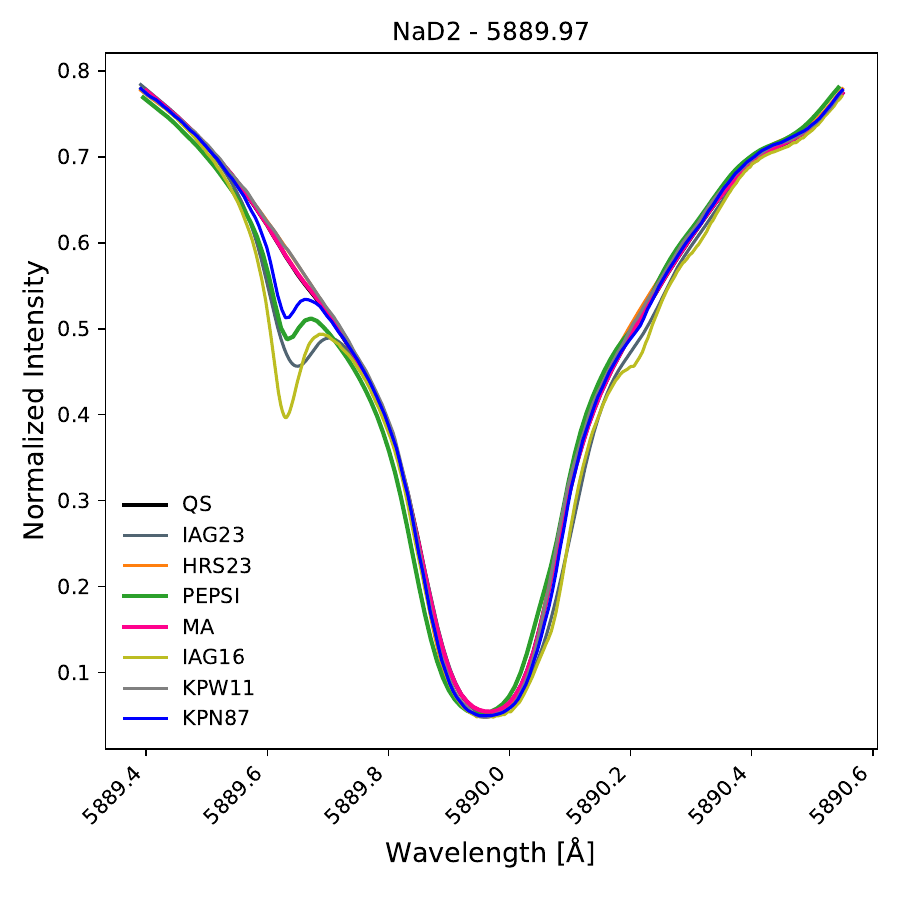}
        \caption{}
        \label{fig:NaD2}
    \end{subfigure}
    \hfill
    \begin{subfigure}[b]{0.45\linewidth}
        \centering
        \includegraphics[width=\textwidth, height=0.8\textwidth, keepaspectratio]{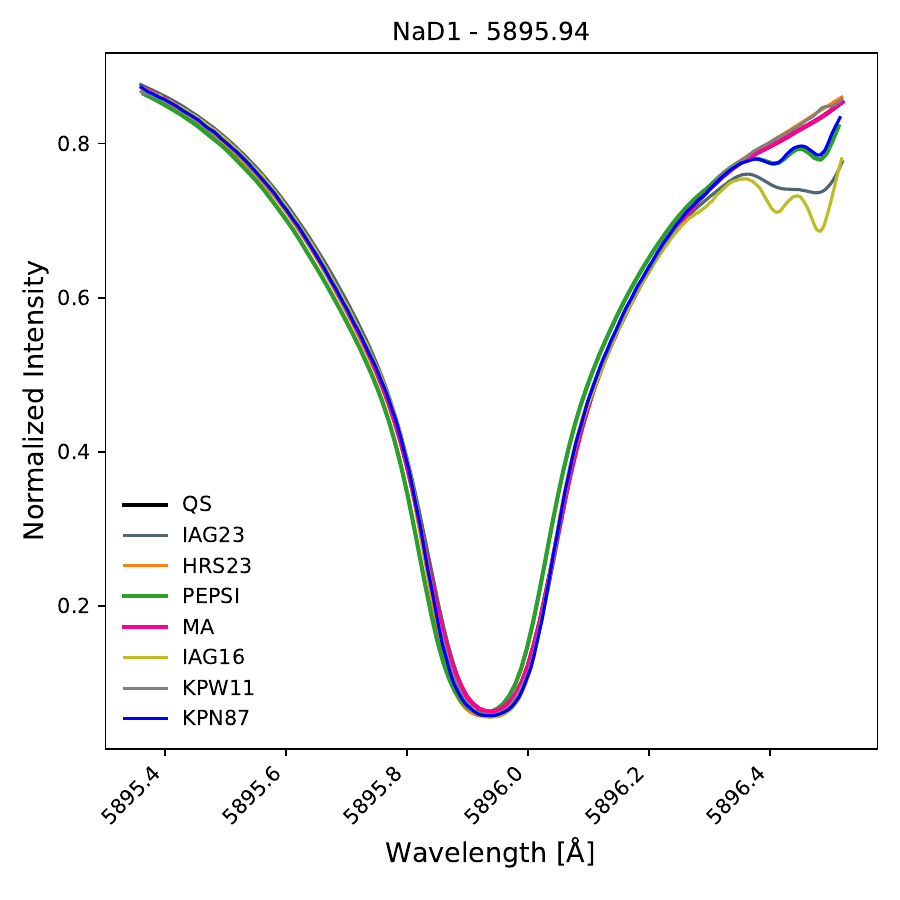}
        \caption{}
        \label{fig:NaD1}
    \end{subfigure}
    \hfill
    \begin{subfigure}[b]{0.45\linewidth}
        \centering
        \includegraphics[width=\textwidth, height=0.8\textwidth, keepaspectratio]{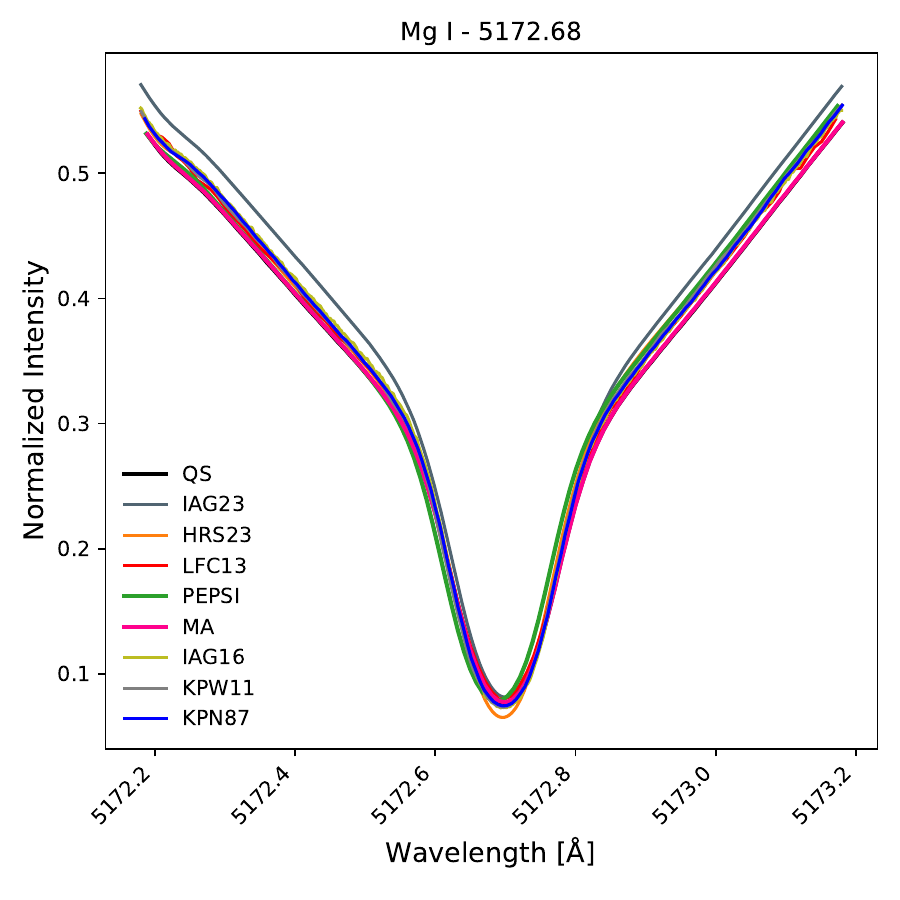}
        \caption{}
        \label{fig:Mg}
    \end{subfigure}
    \hfill
    \begin{subfigure}[b]{0.45\linewidth}
        \centering
        \includegraphics[width=\textwidth, height=0.8\textwidth, keepaspectratio]{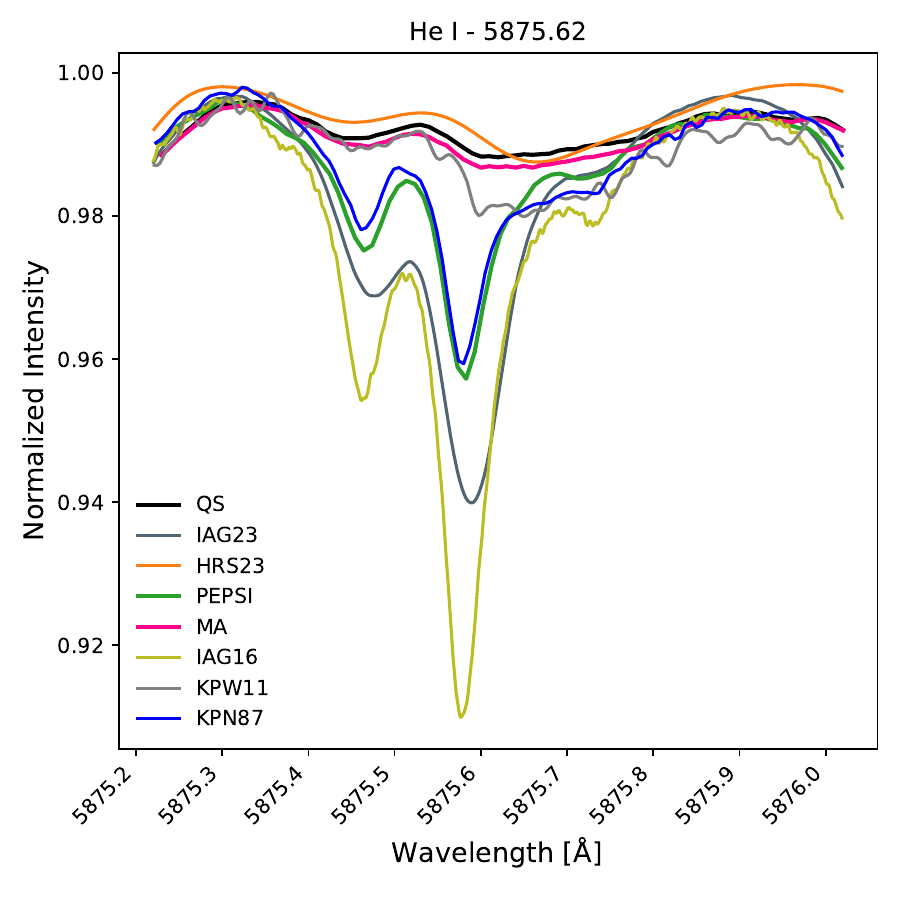}
        \caption{}
        \label{fig:He}
    \end{subfigure}
    \hfill
    \begin{subfigure}[b]{0.45\linewidth}
        \centering
        \includegraphics[width=\textwidth]{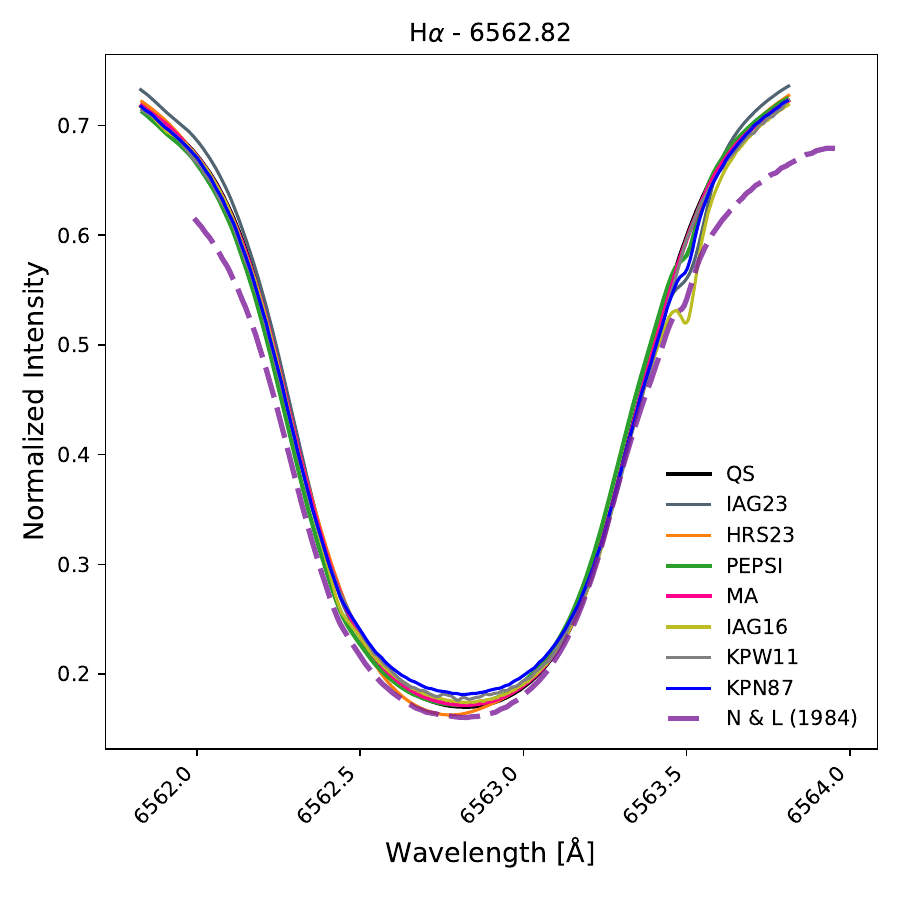}
        \caption{}
        \label{fig:HalphaComparison}
    \end{subfigure}
    \hfill
    \begin{subfigure}[b]{0.45\linewidth}
        \centering
        \includegraphics[width=\textwidth]{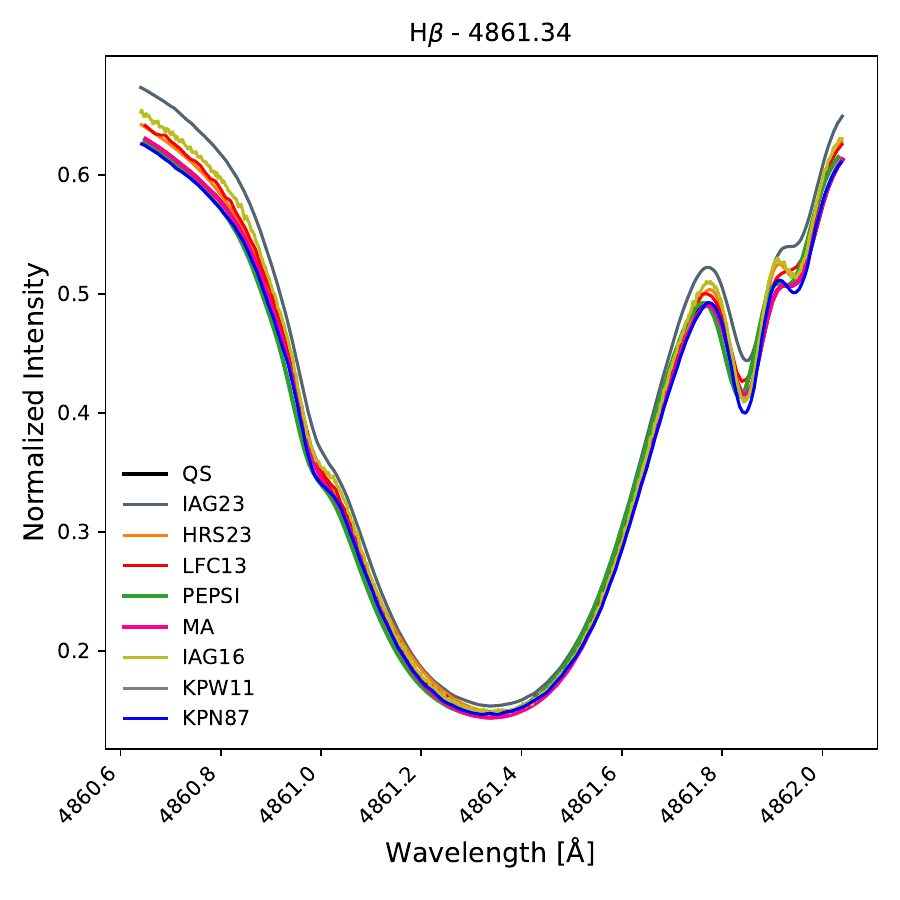}
        \caption{}
        \label{fig:Hbeta}
    \end{subfigure}
    \captionsetup{justification=raggedright,singlelinecheck=false} 
    \caption{Illustration of some of the most-investigated broad lines present in the atlases studied in this manuscript. For comparison, we have included the \Halpha{} line from the disc-centre atlas by \citet{Neckel1984SoPh...90..205N} which was obtained from \citet{DiazBaso2021zndo...5608441D}.}
    \label{fig:All_figures}
\end{figure*}

Our results reveal that the narrow spectral lines of other atlases generally exhibit smaller EWs compared to our QS atlas, whereas the broad lines (apart H$\beta$) show the opposite behaviour (see Fig.~\ref{fig:normalised_ews}). We computed EWs as they are widely regarded as metrics independent of spectral resolution and instrumental profiles. However, our findings challenge this assumption. The EWs of our MA atlas were nearly identical to those of the QS atlas, with an average fractional difference of only 0.05\% across all lines. In contrast, the EWs of other atlases showed noticeably larger deviations relative to our QS atlas, with average fractional differences in the range of 0.7–1.2\% for all lines and 0.6–1.6\% for the broad lines alone. The large discrepancies between the EWs of our QS and MA atlases and those of \citetalias{Molaro2013A&A...560A..61M}, which is characterised by the same spectral resolution, suggest that instrumentation and differing observational conditions have a dominant influence on the EWs of spectral lines in an atlas than solar activity.

We also introduce the activity number (Section~\ref{subsec: an}) as a novel metric for solar activity. This metric is computed as the difference of ratios of other atlases relative to QS and MA reference atlases. To minimise instrumental effects, we degraded the resolution of each spectral line uniquely to best match the QS atlas. Instead of applying the same convolution to all spectral lines of a given atlas, we chose to use a tailored convolution for each line that minimises its corresponding activity number. The rationale behind this approach is related to the difficulty in disentangling instrumental effects from intrinsic spectral features, where applying a uniform convolution could lead to artificially inflated activity values; such values may be misinterpreted by readers as absolute indicators of solar activity. By selecting the convolution that minimises the activity number for each line, this method is aimed at avoiding potential misinterpretation of our results. Despite this improvement, instrumental differences still appeared to be prominent, particularly for broad lines (see Fig.~\ref{fig:All_ANs}).  

Both EWs and activity numbers showed strong general correlations with average core formation height. Since the instrumental and observational differences between the QS and MA atlases are minimal compared to those with other atlases, the strong linear trend observed in the activity numbers of our MA atlas is particularly noteworthy (see Fig.~\ref{fig:AN_per_atlas}). Whether a similar linear trend can be observed for active atlases compared to quiet Sun atlases using other instruments remains to be seen, making this an interesting avenue for future investigations.

Additionally, we sought to gain insight into the specific RVs of single spectral lines to enable a further comparison of the atlases. The RVs of the MA atlas differ by only about 10\,m\,s$^{-1}$, whereas the other atlases exhibit RV differences of the order of hundreds of m\,s$^{-1}$. These findings further support the conclusion that differences between the atlases are primarily driven by instrumental effects and varying observational conditions rather than solar activity. To achieve more conclusive results, an investigation involving hundreds of spectral lines would be recommended.

The \CaIIHK\ lines were excluded from our EW and activity number calculations due to the large intensity differences between our QS and MA atlases compared to other atlases that also cover this spectral region. As discussed in Section~\ref{sec:CaLines}, the results of assumptions in the normalisation techniques used to construct the early Kitt Peak atlases have been propagated by many subsequent studies. Comparisons with synthetic spectra (see Fig.~\ref{fig:SyntheticCaLines}) revealed a closer match with our QS atlas, which was constructed using synthetic spectra from the POLLUX database, compared to \citetalias{neckel1987spectral}. Given the ubiquity of the \CaIIHK\ lines in solar and stellar physics, these significant intensity differences warrant further detailed investigation to fully understand their impact.

An absolute benchmark atlas representing quiet Sun conditions is essential for accurate solar and stellar analysis. Our QS atlas, derived as the median of the 162 quietest days of solar cycle 24 with minimal contamination, serves as such a reference, benefiting from effective telluric correction and an average spot number of 0 (see Table~\ref{tab:spot numbers}). Atlases capturing higher activity levels are equally valuable, as comparisons with the reference atlas help reveal the impact of solar activity on the spectrum. However, our findings suggest that instrumental effects dominate over activity-related variations, emphasising the importance of constructing quiet Sun references using the same instrument for precision work. Even though differences in spectral resolution, Doppler shifts, and intensity shifts introduced by line convolution present challenges for comparing atlases observed with different instruments, an optimisation model could address these discrepancies by determining the best parameters for matching spectral lines to a reference atlas. In this context, our new activity number approach shows promise as a reliable metric capable of calculating absolute solar activity values, along with tracking time series variations.

\begin{acknowledgements}
This work is the result of an internship led by AP and MD
We acknowledge Paris Observatory for the use of their spectroheliograms, and Specola Solare Ticinese for use of their sunspot drawings.
We acknowledge productive discussions with Dr Krishna Mooroogen.
We are very grateful to Nicholas Galateo for sharing his Python expertise which greatly aided the development of the code used in this work. We acknowledge Dr. Jeffrey Hall for timely and helpful response to our request for information about his paper \citet{Hall2007}.
AP and MD are supported by the European Research Council (ERC) under the European Union’s Horizon 2020 research and innovation program (grant agreement No. 833251 PROMINENT ERC-ADG 2018).
AP is supported by the \emph{Deut\-sche For\-schungs\-ge\-mein\-schaft, DFG\/} project number PI 2102/1-1
M.E. acknowledges financial support through the SPP1992 under Project DFG RE 1664/17-1.
This research has made use of NASA's Astrophysics Data System (ADS) bibliographic services. 
We acknowledge the community efforts devoted to the development of the following open-source packages that were used in this work: numpy (\href{http:\\numpy.org}{numpy.org}), matplotlib (\href{http:\\matplotlib.org}{matplotlib.org}), astropy (\href{http:\\astropy.org}{astropy.org}), scipy (\href{https://scipy.org/}{scipy.org}), and scikit-learn (\href{https://scikit-learn.org/stable/#}{scikit-learn.org}).
We extensively used the ISPy library \citep{ISPy2021}, and SOAImage DS9 \citep{2003DS9} for data visualisation. 
This work has made use of the VALD database, operated at Uppsala University, the Institute of Astronomy RAS in Moscow, and the University of Vienna.
\end{acknowledgements}

\bibliographystyle{aa}
\bibliography{refbib}

\begin{appendix}
\renewcommand{\thesection}{A} 
\section{Activity number} \label{appendix:activity_number}
\renewcommand{\theequation}{A\arabic{equation}} 
\setcounter{equation}{0} 

We note some caveats that should be considered when using this metric to measure activity. Firstly that this definition relies on the two reference atlases ideally capturing 'activity' across the spectrum. The reality is certainly more complex, with different types of activity (filament, plage, spot, etc.) potentially increasing or decreasing intensities in some wavelengths. Further, at any intersection points where, at a given wavelength, the two reference atlases have equal intensities any differences in intensity are considered irrelevant for measuring activity.

Here, we provide a more detailed description of our new activity number metric by inspecting its properties in various scenarios to aid with its interpretation. The first term in Equation.~\ref{eq:activity_number}, ($D_{A}$), can be re-written in the form

\begin{equation}
    D_{A} = I_{A}\left(\frac{I_{MA}-I_{QS}}{I_{QS}I_{MA}}\right),
\label{eq:B1}
\end{equation}
\begin{itemize}
    \item at points where $I_{MA} > I_{QS}$     (activity enhances emission), $D_{A}$ is positive;
    \item at points where $I_{MA} < I_{QS}$     (activity decreases emission), $D_{A}$ is negative;
    \item at points where $I_{MA} = I_{QS}$     (activity is irrelevant), $D_{A}$ is zero.
\end{itemize}
Therefore, we are defining a quantity that is positive where activity is associated with increased emission, negative where activity is associated with decreased emission and zero where intensity is modelled as not being impacted by solar activity.

When comparing atlases, the difference due to QS (minimum activity reference atlas) must also be subtracted, as we have done in this paper. The activity number metric, from Equation.~\ref{eq:activity_number} then becomes

\begin{equation}
    D_{A,QS} = I_{A}\left(\frac{I_{MA}-I_{QS}}{I_{QS}I_{MA}}\right) - I_{QS}\left(\frac{I_{MA}-I_{QS}}{I_{QS}I_{MA}}\right),
\label{eq:B2}
\end{equation}

\begin{equation}
    D_{A,QS} = \left(I_{A} - I_{QS}\right)\left(\frac{I_{MA}-I_{QS}}{I_{QS}I_{MA}}\right).
\label{eq:B3}
\end{equation}

There are five scenarios or types of atlases that we  considered within this metric, listed below.

\begin{enumerate}

\item For a high activity atlas with intensities, $I_{A}$:

\begin{itemize}
    \item in wavelengths where activity enhances emission, $I_{A} > I_{MA} > I_{QS}$,
$D_{A,QS}= (+)(+) >0$;
    \item in wavelengths where activity decreases emission, $I_{A} < I_{MA} < I_{QS}$,
$D_{A,QS}= (-)(-) >0$;
    \item at points where activity is considered irrelevant, $I_{QS} = I_{MA}$, $D_{A,QS}$ is zero.
\end{itemize}

\item For an active atlas with intensities $I_{A}=I_{MA}$, Equation.~\ref{eq:B3} becomes $D_{A,QS} = \frac{(I_{MA} - I_{QS})^2}{I_{QS} I_{MA}}$ and therefore: 

\begin{itemize}
    \item in wavelengths where activity enhances emission, $I_{A} = I_{MA} > I_{QS}$,
$D_{A,QS}= (+) >0$;
    \item in wavelengths where activity decreases emission. $I_{A} = I_{MA} < I_{QS}$,
$D_{A,QS}= (+) >0$;
    \item at points where activity is considered irrelevant, $I_{QS} = I_{MA}$, $D_{A,QS}$ is zero.
\end{itemize}

\item For a moderate activity atlas, with intensities $I_{A}$, which is more active than the QS atlas, but less than the MA atlas: 

\begin{itemize}
    \item in wavelengths where activity enhances emission, $I_{MA} > I_{A} > I_{QS}$,
$D_{A,QS}= (+)(+) >0$;
    \item in wavelengths where activity decreases emission, $I_{MA} < I_{A} < I_{QS}$,
$D_{A,QS}= (-)(-) >0$;
    \item at points where activity is considered irrelevant, $I_{QS} = I_{MA}$, $D_{A,QS}$ is zero.
\end{itemize}

\item For a minimum activity atlas with intensities $I_{A}=I_{QS}$ in all wavelengths, the activity number is zero in all wavelengths. 

\item For the hypothetical case of having a low activity atlas, with intensities $I_{A}$, which is less active than the referenced QS atlas:

\begin{itemize}
    \item in wavelengths where activity enhances emission, $I_{MA} > I_{QS} > I_{A}$,
$D_{A,QS}= (-)(+) <0$;
    \item in wavelengths where activity decreases emission, $I_{MA} < I_{QS} < I_{A}$,
$D_{A,QS}= (+)(-) <0$;
    \item at points where activity is considered irrelevant, $I_{QS} = I_{MA}$, $D_{A,QS}$ is zero.
\end{itemize}

\end{enumerate}

\onecolumn 
\newpage
\renewcommand{\thesection}{B} 
\addcontentsline{toc}{section}{Appendix B}

\section{QS and MA spot numbers}
\renewcommand{\thetable}{B\arabic{table}}
\setcounter{table}{0} 

\captionsetup{type=table}
\begin{center}
    \captionsetup{justification=raggedright,singlelinecheck=false}
    \caption{Observation dates for our QS atlas and their respective spot numbers.}
    \begin{tabular}{|c|c|c|c|c|c|c|c|c|c|c|c|}
    \hline
        date & \multicolumn{1}{p{0.6cm}|}{\centering Spot \\ Nr} & date & \multicolumn{1}{p{0.6cm}|}{\centering Spot \\ Nr} & date & \multicolumn{1}{p{0.6cm}|}{\centering Spot \\ Nr} & date & \multicolumn{1}{p{0.6cm}|}{\centering Spot \\ Nr} & date & \multicolumn{1}{p{0.6cm}|}{\centering Spot \\ Nr} & date & \multicolumn{1}{p{0.6cm}|}{\centering Spot \\ Nr} \\ \hline
22/02/2018 & 0 & 04/11/2018 & 0 & 26/04/2019 & 0 & 12/09/2019 & 0 & 01/12/2019 & 0 & 22/02/2020 & 0 \\ \hline
20/03/2018 & 0 & 05/11/2018 & 0 & 27/04/2019 & 0 & 13/09/2019 & 0 & 03/12/2019 & 0 & 24/02/2020 & 0 \\ \hline
21/03/2018 & 0 & 06/11/2018 & 0 & 28/04/2019 & 0 & 14/09/2019 & 0 & 04/12/2019 & 0 & 26/02/2020 & 0 \\ \hline
22/03/2018 & 0 & 07/11/2018 & 0 & 29/04/2019 & 0 & 15/09/2019 & 0 & 05/12/2019 & 0 & 27/02/2020 & 0 \\ \hline
23/03/2018 & 0 & 29/11/2018 & 0 & 30/04/2019 & 0 & 18/09/2019 & 0 & 09/12/2019 & 0 & 28/02/2020 & 0 \\ \hline
24/03/2018 & 0 & 30/11/2018 & 0 & 01/05/2019 & 0 & 19/09/2019 & 0 & 10/12/2019 & 0 & 29/02/2020 & 0 \\ \hline
25/03/2018 & 0 & 01/12/2018 & 0 & 02/05/2019 & 0 & 20/09/2019 & 0 & 12/12/2019 & 0 & 01/03/2020 & 0 \\ \hline
09/04/2018 & 0 & 02/12/2018 & 0 & 03/05/2019 & 0 & 21/09/2019 & 0 & 13/12/2019 & 0 & 04/03/2020 & 0 \\ \hline
02/05/2018 & 0 & 03/12/2018 & 0 & 20/05/2019 & 0 & 22/09/2019 & 0 & 14/12/2019 & 0 & 22/03/2020 & 0 \\ \hline
03/07/2018 & 0 & 04/12/2018 & 0 & 21/05/2019 & 0 & 23/09/2019 & 0 & 15/12/2019 & 0 & 25/03/2020 & 0 \\ \hline
04/07/2018 & 0 & 09/12/2018 & 0 & 22/05/2019 & 0 & 26/09/2019 & 0 & 16/12/2019 & 0 & 12/04/2020 & 0 \\ \hline
05/07/2018 & 0 & 24/12/2018 & 0 & 23/05/2019 & 0 & 27/09/2019 & 0 & 19/12/2019 & 0 & 14/04/2020 & 0 \\ \hline
06/07/2018 & 0 & 11/01/2019 & 0 & 24/05/2019 & 0 & 28/09/2019 & 0 & 21/12/2019 & 0 & 20/04/2020 & 0 \\ \hline
07/07/2018 & 0 & 12/01/2019 & 0 & 18/06/2019 & 0 & 29/09/2019 & 0 & 17/01/2020 & 0 & 22/04/2020 & 0 \\ \hline
29/07/2018 & 0 & 13/01/2019 & 0 & 19/06/2019 & 0 & 16/10/2019 & 0 & 18/01/2020 & 0 & 09/05/2020 & 0 \\ \hline
29/08/2018 & 0 & 14/01/2019 & 0 & 20/06/2019 & 0 & 18/10/2019 & 0 & 21/01/2020 & 0 & 11/05/2020 & 0 \\ \hline
30/08/2018 & 0 & 16/01/2019 & 0 & 21/06/2019 & 0 & 20/10/2019 & 0 & 03/02/2020 & 0 & 12/05/2020 & 0 \\ \hline
26/09/2018 & 0 & 19/01/2019 & 0 & 22/06/2019 & 0 & 21/10/2019 & 0 & 06/02/2020 & 0 & 13/05/2020 & 0 \\ \hline
27/09/2018 & 0 & 27/02/2019 & 0 & 27/06/2019 & 0 & 31/10/2019 & 0 & 07/02/2020 & 0 & 14/05/2020 & 0 \\ \hline
28/09/2018 & 0 & 28/02/2019 & 0 & 09/08/2019 & 0 & 08/11/2019 & 0 & 09/02/2020 & 0 & 15/05/2020 & 0 \\ \hline
09/10/2018 & 0 & 01/03/2019 & 0 & 13/08/2019 & 0 & 09/11/2019 & 0 & 10/02/2020 & 0 & 16/05/2020 & 0 \\ \hline
27/10/2018 & 0 & 02/03/2019 & 0 & 14/08/2019 & 0 & 23/11/2019 & 0 & 12/02/2020 & 0 & 11/07/2020 & 0 \\ \hline
29/10/2018 & 0 & 28/03/2019 & 0 & 15/08/2019 & 0 & 24/11/2019 & 0 & 14/02/2020 & 0 & 13/07/2020 & 0 \\ \hline
30/10/2018 & 0 & 22/04/2019 & 0 & 16/08/2019 & 0 & 25/11/2019 & 0 & 17/02/2020 & 0 & 09/09/2020 & 0 \\ \hline
31/10/2018 & 0 & 23/04/2019 & 0 & 18/08/2019 & 0 & 27/11/2019 & 0 & 19/02/2020 & 0 & 21/09/2020 & 0 \\ \hline
01/11/2018 & 0 & 24/04/2019 & 0 & 10/09/2019 & 0 & 28/11/2019 & 0 & 20/02/2020 & 0 & 05/10/2020 & 0 \\ \hline
03/11/2018 & 0 & 25/04/2019 & 0 & 11/09/2019 & 0 & 30/11/2019 & 0 & 21/02/2020 & 0 & 06/10/2020 & 0 \\ \hline
    \end{tabular}
    \label{tab:QS_Spot_Numbers}
\end{center}

\begin{table}[H]
    \centering
    \captionsetup{justification=raggedright,singlelinecheck=false}
    \caption{Observation dates for our MA atlas and their respective spot numbers.}
    \begin{tabular}{|c|c|c|c|c|c|c|c|c|c|c|c|}
    \hline
        date & \multicolumn{1}{p{0.6cm}|}{\centering Spot \\ Nr} & date & \multicolumn{1}{p{0.6cm}|}{\centering Spot \\ Nr} & date & \multicolumn{1}{p{0.6cm}|}{\centering Spot \\ Nr} & date & \multicolumn{1}{p{0.6cm}|}{\centering Spot \\ Nr} & date & \multicolumn{1}{p{0.6cm}|}{\centering Spot \\ Nr} & date & \multicolumn{1}{p{0.6cm}|}{\centering Spot \\ Nr} \\ \hline
17/07/2015 & 50 & 19/09/2015 & 64 & 29/11/2015 & 51 & 28/01/2016 & 77 & 16/04/2016 & 42 & 13/08/2016 & 58 \\ \hline
20/07/2015 & 36 & 21/09/2015 & 78 & 30/11/2015 & 47 & 30/01/2016 & 36 & 28/04/2016 & 90 & 14/08/2016 & 59 \\ \hline
30/07/2015 & 76 & 22/09/2015 & 86 & 12/12/2015 & 83 & 01/02/2016 & 44 & 29/04/2016 & 82 & 15/08/2016 & 63 \\ \hline
31/07/2015 & 69 & 24/09/2015 & 101 & 13/12/2015 & 81 & 04/02/2016 & 108 & 30/04/2016 & 89 & 30/08/2016 & 74 \\ \hline
01/08/2015 & 64 & 28/09/2015 & 156 & 16/12/2015 & 63 & 06/02/2016 & 79 & 01/05/2016 & 87 & 31/08/2016 & 77 \\ \hline
02/08/2015 & 53 & 29/09/2015 & 120 & 22/12/2015 & 72 & 09/02/2016 & 84 & 02/05/2016 & 83 & 01/09/2016 & 78 \\ \hline
03/08/2015 & 61 & 30/09/2015 & 95 & 23/12/2015 & 68 & 10/02/2016 & 75 & 06/05/2016 & 57 & 06/09/2016 & 43 \\ \hline
04/08/2015 & 90 & 20/10/2015 & 95 & 24/12/2015 & 69 & 11/02/2016 & 85 & 14/05/2016 & 88 & 07/09/2016 & 57 \\ \hline
05/08/2015 & 94 & 27/10/2015 & 67 & 25/12/2015 & 76 & 12/02/2016 & 63 & 15/05/2016 & 87 & 08/09/2016 & 53 \\ \hline
06/08/2015 & 108 & 28/10/2015 & 81 & 26/12/2015 & 65 & 13/02/2016 & 38 & 16/05/2016 & 55 & 09/09/2016 & 79 \\ \hline
08/08/2015 & 93 & 29/10/2015 & 93 & 28/12/2015 & 71 & 14/02/2016 & 46 & 17/05/2016 & 38 & 10/09/2016 & 78 \\ \hline
21/08/2015 & 73 & 30/10/2015 & 88 & 29/12/2015 & 57 & 15/02/2016 & 50 & 18/05/2016 & 30 & 11/09/2016 & 67 \\ \hline
22/08/2015 & 71 & 31/10/2015 & 86 & 30/12/2015 & 37 & 16/02/2016 & 42 & 19/05/2016 & 43 & 05/10/2016 & 41 \\ \hline
23/08/2015 & 81 & 02/11/2015 & 101 & 31/12/2015 & 22 & 17/02/2016 & 38 & 21/05/2016 & 18 & 07/10/2016 & 60 \\ \hline
26/08/2015 & 47 & 03/11/2015 & 82 & 01/01/2016 & 37 & 22/02/2016 & 34 & 23/05/2016 & 16 & 08/10/2016 & 63 \\ \hline
27/08/2015 & 55 & 04/11/2015 & 100 & 07/01/2016 & 61 & 23/02/2016 & 37 & 17/06/2016 & 39 & 10/10/2016 & 66 \\ \hline
28/08/2015 & 44 & 05/11/2015 & 82 & 09/01/2016 & 97 & 02/03/2016 & 78 & 18/06/2016 & 47 & 11/10/2016 & 63 \\ \hline
29/08/2015 & 56 & 07/11/2015 & 82 & 10/01/2016 & 90 & 03/03/2016 & 77 & 10/07/2016 & 50 & 12/10/2016 & 41 \\ \hline
30/08/2015 & 56 & 08/11/2015 & 75 & 11/01/2016 & 47 & 04/03/2016 & 111 & 11/07/2016 & 62 & 02/04/2017 & 83 \\ \hline
31/08/2015 & 35 & 09/11/2015 & 66 & 12/01/2016 & 40 & 05/03/2016 & 83 & 12/07/2016 & 59 & 03/04/2017 & 100 \\ \hline
01/09/2015 & 43 & 11/11/2015 & 61 & 13/01/2016 & 38 & 06/03/2016 & 78 & 13/07/2016 & 53 & 04/04/2017 & 76 \\ \hline
02/09/2015 & 36 & 15/11/2015 & 56 & 17/01/2016 & 58 & 07/03/2016 & 79 & 14/07/2016 & 58 & 03/09/2017 & 105 \\ \hline
03/09/2015 & 29 & 16/11/2015 & 38 & 18/01/2016 & 54 & 08/03/2016 & 63 & 15/07/2016 & 69 & 04/09/2017 & 112 \\ \hline
04/09/2015 & 37 & 19/11/2015 & 41 & 19/01/2016 & 66 & 09/03/2016 & 91 & 16/07/2016 & 60 & 05/09/2017 & 119 \\ \hline
11/09/2015 & 81 & 21/11/2015 & 58 & 20/01/2016 & 62 & 10/03/2016 & 80 & 17/07/2016 & 39 & 06/09/2017 & 100 \\ \hline
12/09/2015 & 89 & 22/11/2015 & 71 & 21/01/2016 & 64 & 11/03/2016 & 51 & 18/07/2016 & 64 & 07/09/2017 & 97 \\ \hline
13/09/2015 & 83 & 24/11/2015 & 59 & 22/01/2016 & 67 & 12/03/2016 & 70 & 19/07/2016 & 59 & 08/09/2017 & 88 \\ \hline
15/09/2015 & 68 & 25/11/2015 & 61 & 24/01/2016 & 48 & 16/03/2016 & 79 & 20/07/2016 & 56 &  &  \\ \hline
16/09/2015 & 86 & 26/11/2015 & 65 & 25/01/2016 & 67 & 18/03/2016 & 47 & 08/08/2016 & 81 & &  \\ \hline
17/09/2015 & 75 & 27/11/2015 & 61 & 26/01/2016 & 69 & 14/04/2016 & 46 & 10/08/2016 & 77 &  &  \\ \hline
18/09/2015 & 71 & 28/11/2015 & 51 & 27/01/2016 & 83 & 15/04/2016 & 33 & 12/08/2016 & 71 &  &  \\ \hline
    \end{tabular}
    \label{tab:MA_Spot_Numbers}
\end{table}

\end{appendix}

\end{document}